\documentclass[12pt]{iopart}
\usepackage{graphicx}
\eqnobysec

\newcommand{\beq}{\begin{equation}}
\newcommand{\beqa}{\begin{eqnarray}}
\newcommand{\eeq}{\end{equation}}
\newcommand{\eeqa}{\end{eqnarray}}
\newcommand{\abs}[1]{\vert#1\vert}
\newcommand{\bigmean}[1]{\left\langle#1\right\rangle}
\renewcommand{\d}{{\rm d}}

\newcommand{\dmean}[1]{\langle\!\langle#1\rangle\!\rangle}
\newcommand{\dpar}{\partial}
\newcommand{\diff}[1]{\frac{\dpar}{\dpar#1}}
\newcommand{\ds}{\displaystyle}
\renewcommand{\e}{{\rm e}}
\newcommand{\eps}{\varepsilon}
\newcommand{\frad}[2]{\ds{\frac{\ds#1}{\ds#2}}}
\renewcommand{\i}{{\rm i}}
\newcommand{\ito}{{\rm I}}
\newcommand{\iito}{[\ito]\qquad}
\renewcommand{\max}{{\rm max}}
\newcommand{\mean}[1]{\langle#1\rangle}
\renewcommand{\min}{{\rm min}}
\newcommand{\str}{{\rm S}}
\newcommand{\sstr}{[\str]\qquad}
\renewcommand{\sup}[1]{^{(#1)}}
\newcommand{\var}{\mathop{\rm var}\nolimits}
\newcommand{\xxi}{{\bar{\xi}}}
\newcommand{\w}{\tilde}
\newcommand{\zed}{{\cal{Z}}}

\renewcommand{\Re}{\mathop{\rm Re}\nolimits}
\renewcommand{\S}{\Sigma}
\newcommand{\Z}{{\bar{Z}}}

\begin{document}

\title{On an imaginary exponential functional of Brownian motion}

\author{D Gredat$^1$, I Dornic$^1$, and J M Luck$^2$}

\address{$^1$ Service de Physique de l'Etat Condens\'e, IRAMIS/SPEC, CEA Saclay
and URA~2464, CNRS, 91191 Gif-sur-Yvette cedex, France}

\address{$^2$ Institut de Physique Th\'eorique, IPhT, CEA Saclay
and URA~2306, CNRS, 91191 Gif-sur-Yvette cedex, France}

\begin{abstract}
We investigate a random integral which provides a natural example
of an imaginary exponential functional of Brownian motion.
This functional shows up in the study
of the binary annihilation process,
within the Doi-Peliti formalism for reaction-diffusion systems.
The main emphasis is put on the complementarity between
the usual Langevin approach and another approach based on the similarity
with Kesten variables and other one-dimensional disordered systems.
Even though neither of these routes leads to the full solution of the problem,
we have obtained a collection of results describing various regimes of interest.
\end{abstract}


\eads{\mailto{damien.gredat@cea.fr},\mailto{ivan.dornic@cea.fr},\mailto{jean-marc.luck@cea.fr}}

\maketitle

\section{Introduction}
\label{intro}

Real exponential functionals of Brownian motion
have been the subject of much activity in probability theory~\cite{yor}.
They have found many applications to problems ranging from finance to
physics~\cite{comtet}.
To be more specific, if $B(t)$ is Brownian motion
(a Wiener process with $B(0)=0$ and $\mean{B(t)^2}=t$),
the following stochastic variable
\beq
X(t)=\int_0^t\e^{-s+gB(s)}\d s,
\label{Xtdef}
\eeq
where $g$ is a real coupling constant,
shows up in one guise or another in various models
of disordered systems~\cite{bouchaud,opper}.
The above random integral can be shown to represent the solution
of the following Langevin equation with multiplicative noise:
\beq
\frac{\d X(t)}{\d t}=1-X(t)+gX(t)\eta(t),
\label{xintro}
\eeq
with $\eta(t)\equiv\d B(t)/\d t$ being a zero-mean Gaussian white noise,
normalized as
\beq
\mean{\eta(s)\eta(t)}=\delta(t-s).
\label{etanor}
\eeq
The derivation of the solution to~(\ref{xintro}) in the form~(\ref{Xtdef}),
including the required stochastic calculus prescription (Stratonovich),
will be reviewed in Section~\ref{real}.
As an illustration of the versatility of the
situations in which the random variable $X(t)$ appears,
the Langevin equation~(\ref{xintro}) may serve to study
the effect of multiplicative noise on a deterministic fixed point.
Indeed, the linear force $1-X(t)$ stabilizes $X(t)$
at the fixed point $X=1$ in the absence of noise.
The solution $X(t)$ however keeps on fluctuating forever
because of the multiplicative noise term $gX(t)\eta(t)$.
In the long-time limit, what emerges out of this interplay between the
deterministic restoring force and the fluctuating noise term
is a non-trivial distribution for the random variable
\beq
X=\int_0^\infty\e^{-t+gB(t)}\d t,
\label{xdef}
\eeq
representing the stationary solution of~(\ref{xintro}).
The distribution of $X\equiv\lim_{t\to\infty}X(t)$
can be computed in a number of ways,
and we shall review in Section~\ref{real} two useful methods to do so.
Using either more elaborate path-integral methods or probabilistic identities,
the full-time dependent distribution of $X(t)$ can also be obtained,
and features, among other things, an interesting continuous spectrum of
relaxation rates~\cite{comtet}.

A seemingly innocuous and somewhat natural generalization of~(\ref{xdef})
consists in analytically continuing the coupling constant $g\to\i g$.
We thus obtain the following random integral,
defining an imaginary exponential functional of Brownian motion:
\beq
Z=\int_0^\infty\e^{-t+\i gB(t)}\d t,
\label{zdef}
\eeq
which now lives in the complex plane (actually in the unit disk).

The main goal of the present article is to investigate the distribution
of the functional $Z$.
Throughout the following, instead of the coupling constant $g$,
we rather make use of the dimensionless diffusion constant
\beq
D=\frac{g^2}{2}.
\label{ddef}
\eeq

As hinted at above, even though a large number of works have been devoted to
real exponential functionals of Brownian motion,
much less is known
about the distribution of complex functionals of Brownian motion
such as~(\ref{zdef}).
We found interesting from a conceptual viewpoint to tackle this problem,
in particular in order to see if any of the methods which proved successful
for the real case would extend to this complex-variable setting.
Besides this,
it turns out that complex stochastic processes have surfaced time and again in
different scientific disciplines
ranging from signal theory, where processes involving imaginary
exponential functionals of Brownian motion
occur in the study of {\it phase noise}~\cite{ie3},
to quantum optics~\cite{gardsmith}, in conjunction with the development of phase-space
representations and of
the associated formalism of {\it quasi-probabilities},
and finally to reaction-diffusion processes~\cite{mun,delou}
through the Doi-Peliti approach~\cite{doipeliti}.
The latter topic has constituted our original thrust to embark
on the study of~(\ref{zdef}).
Let us now describe how the connection emerges.

In the context of interacting particle systems such as reaction-diffusion systems,
there exists a standard set of techniques,
usually referred to as the Doi-Peliti formalism~\cite{doipeliti}
(see~\cite{tauber} for a recent review, and the references therein),
which allows to recast the
master equation describing the evolution of these processes in terms of
a field theory whose action involves a pair of conjugate fields.
Without entering into much detail,
provided certain technical conditions are met,
the theory can in turn be transformed into
a Langevin equation for a single {\it density field} $\varphi_n(t)$,
customarily dubbed so because its noise-average $\mean{\varphi_n(t)}$,
coincides with the local mean particle number for the underlying
reaction-diffusion process $\dmean{\rho_n(t)}$.
Here the double brackets denote an average over the dynamics of the particles.
In the case of the binary annihilation reaction
\beq
\label{RDS}
A+A\to\emptyset,
\eeq
where particles $A$ diffuse by hopping with rate $\kappa$
on a hypercubic lattice in dimension~$d$,
and annihilate pairwise with rate
$\lambda$ when they meet on a given lattice site $n$,
one can show~\cite{mun,delou,doipeliti} that the stochastic density field
$\varphi_n(t)$ obeys the following Langevin-It\^o equation:
\beq
\label{Doilattice}
\frac{\d\varphi_n(t)}{\d t}=\kappa(\nabla^2\varphi)_n(t)-\lambda\varphi_n(t)^2+\zeta_n(t),
\eeq
with $\nabla^2$ being the lattice Laplacian, such that, e.g.,
$(\nabla^2\varphi)_n=\varphi_{n+1}+\varphi_{n-1}-2\varphi_n$ in one dimension.
It can be expected on physical grounds that the amplitude
of the Gaussian noise term $\zeta_n[\varphi_n]$ vanishes when $\varphi_n=0$.
Indeed, the first two terms on the right-hand side of~(\ref{Doilattice})
respectively account for the diffusion of the particles
and for the (mean-field) decay rate of the particle density due to pairwise annihilation.
If the noise amplitude vanishes when $\varphi_n=0$,
there is no evolution at all in regions where $\varphi_n=0$.
The Doi-Peliti approach also yields the following expression
for the correlator of the noise:
\beq
\label{correl}
\mean{\zeta_m(s)\zeta_n(t)}=-\lambda\varphi_n(t)^2\delta_{m,n}\delta(t-s),
\eeq
which therefore has a negative variance.
In other words, we have
$\zeta_n(t)=\i\sqrt{\lambda}\varphi_n(t)\eta_n(t)$,
where $\eta_n(t)$ is a normalized real Gaussian white noise.
This explains why~(\ref{Doilattice}) is often referred to
as an {\it imaginary-noise} equation.
One would obtain the same Langevin equation
using Gardiner's quasi-probability formalism,
where the particle-number probability distribution is represented as
a superposition of Poisson distributions with
weights $\varphi_n$~\cite{delou,gardiner}.
The equivalence between the Doi-Peliti formalism
and Gardiner's Poisson representation method
has been demonstrated in general in~\cite{droz}.
There is no contradiction in either formalism,
as soon as the auxiliary field $\varphi_n(t)$ is complex-valued,
provided one refrains from erroneously identifying $\varphi_n(t)$
with the (integer-valued) stochastic variable $\rho_n(t)$,
based on the sole equality between the mean values $\mean{\varphi_n(t)}=\dmean{\rho_n(t)}$.
In fact, the precise relationship between the distribution of $\varphi_n(t)$
and that of $\rho_n(t)$
is that the {\it ordinary} moments of $\varphi_n(t)$ are equal to the {\it factorial}
moments of $\rho_n(t)$.
This is a particular instance of what is referred to as duality between two stochastic
processes in the probabilistic literature~\cite{liggett}.
Thus in particular:
\beq
\label{mom2}
\dmean{\rho_n(t)(\rho_n(t)-1)}=\mean{\varphi_n(t)^2},
\eeq
and so one can have $\mean{\varphi_n(t)^2}<0$,
while keeping $\var\rho_n(t)=\dmean{\rho_n(t)^2}-\dmean{\rho_n(t)}^2
=\mean{\varphi_n(t)^2}+\mean{\varphi_n(t)}-\mean{\varphi_n(t)}^2\ge0$.
It is also commonly accepted that the complex-valued nature of the
trajectories of the field $\varphi_n(t)$ is needed in order to account
for the importance of fluctuation effects in low spatial dimensions,
resulting in a slower decay for the total density of particles
than what the naive law of mass action would predict
(viz.~$t^{-d/2}$ vs.~$1/t$ in $d<2$
for the reaction~(\ref{RDS})~\cite{tauber,cardy}).

Henceforth, along the lines of~\cite{mun,delou},
we focus onto the single-site problem associated with~(\ref{Doilattice}),
neglecting any spatial dependence.
This simplification will allow us to better understand the role of the excursions
of the field $\varphi(t)$ in the complex plane.
Setting $\kappa=0$, and absorbing the reaction rate $\lambda$ into the time scale,
one ends up with the following Langevin-It\^o equation for a single complex
stochastic variable $\varphi(t)$:
\beq
\label{Doi}
\frac{\d\varphi(t)}{\d t}=-\varphi(t)^2+\i\varphi(t)\eta(t),
\eeq
where $\eta(t)$ is a normalized Gaussian white noise~(see~(\ref{etanor})).
The initial condition $\varphi(0)$ is real and non-negative
(e.g., $\varphi(0)=\rho_0$ if one starts from a Poisson distribution
with density~$\rho_0$ for the original particle system).
The conjugation symmetry $\varphi\to\bar\varphi$
ensures that the imaginary part of $\varphi(t)$ averages over to zero,
so as to maintain the reality and the non-negativity of
$\mean{\varphi(t)}=\dmean{\rho(t)}$.
It has already been noticed by several authors~\cite{gardsmith,mun,delou}
that~(\ref{Doi}) becomes linear in the variable $\zed(t)=1/\varphi(t)$.
One thus obtains the explicit solution
\beq
\label{Phisol}
\zed(t)=\zed(0)\,\e^{-t/2-\i B(t)}+\int_0^t\e^{-(t-s)/2+\i[B(s)-B(t)]}\,\d s.
\eeq
Rescaling time,
and using the scaling property $B(at)\equiv\sqrt{a}B(t)$ of Brownian motion,
we obtain by identifying~(\ref{zdef}) and~(\ref{Phisol})
the following identity
between the stationary solution $\zed\equiv\lim_{t\to\infty}\zed(t)$
and the functional $Z$ for $g=\sqrt{2}$, i.e., $D=1$~\cite{delou}:
\beq
\label{PhiZ}
\zed=\lim_{t\to\infty}\zed(t)\equiv
2Z\big{|}_{D=1}=2\int_0^\infty\e^{-t+\i\sqrt{2}B(t)}\d t.
\eeq

The above representation of the stationary solution has striking consequences.
At the level of the original process~(\ref{RDS}),
the stationary state is rather featureless,
as there just remains either zero or one particle,
depending on the parity of the initial condition.
By~(\ref{mom2}),
and the corresponding equations for higher-order moments,
this implies that $\mean{\varphi^p}=0$ for any integer $p\ge2$.
Owing to~(\ref{PhiZ}), this reads $\mean{Z^{-p}}\big{|}_{D=1}=0$.
The full distribution of $Z$ will however turn out to be highly non-trivial.
In particular, in stark contrast to what
intuition backing up~(\ref{Doilattice}) or~(\ref{Doi}) could let us foresee,
in the stationary state the variable $\varphi=2/Z$
never reaches the value $\varphi=0$, characteristic of the absorbing state.
Indeed, as already announced, $Z$ lies within the unit disk.

The setup of this article is the following.
In Section~\ref{real}
we present a self-contained investigation of the real exponential functional
$X$~(see~(\ref{xdef})).
The emphasis is put on the complementarity between
the usual Langevin approach and another approach based on the similarity
with the random recursions met in the study of Kesten variables
and one-dimensional disordered systems.
The main section of the paper (Section~\ref{imag}) is devoted to a detailed study
of the imaginary exponential functional $Z$~(see~(\ref{zdef})).
We present numerical illustrations of the distribution of $Z$
and investigate many facets of the problem by analytical means,
including the relationship between the Langevin and Kesten approaches,
the moments $\mean{Z^k\Z^l}$,
the weak-disorder and strong-disorder regimes,
and the asymptotic behavior of the distribution near the unit circle.
Section~\ref{discussion} contains a brief summary of our findings.

\section{A warming up: the real functional}
\label{real}

This section is to a large extent intended as a warming up.
It is devoted to a self-contained study of the real exponential functional
(see~(\ref{xdef}))
\beq
X=\int_0^\infty\e^{-t+gB(t)}\d t.
\label{xdef2}
\eeq
The positive random variable thus defined
is one of the exponential functionals of Brownian motion
which have been investigated in probability theory,
chiefly by Yor and his collaborators (see~\cite{yor} for a review).
It also appears in the physics literature,
in the context of one-dimensional disordered systems~\cite{comtet,bouchaud,opper}.
The coupling constant $g$ measuring the strength of noise,
or, equivalently, the diffusion constant $D$ (see~(\ref{ddef})),
is the sole parameter entering the definition of $X$.

\subsection{Langevin approach}
\label{reallan}

A first approach to study the random variable $X$ consists in using
Langevin equations.
At this point it is useful to recall some elements of
stochastic calculus~\cite{gardiner,risken,kampen}.
A~Langevin equation of the form
\beq
\frac{\d X(t)}{\d t}=a(X(t))+b(X(t))\eta(t)
\eeq
is ambiguous as soon as it is non-linear,
in the sense that the noise $\eta(t)$ multiplies a non-trivial function
$b(X(t))$ of the position $X(t)$.
This ambiguity due to the usage of a continuous-time formalism
can be lifted in many ways.
The two most useful and well-known prescriptions are the following
(see~\cite{gardiner,kampen} for a detailed exposition):

\begin{itemize}

\item {\it Stratonovich prescription.}
The Langevin-Stratonovich differential equation
\beq
\sstr\frac{\d X(t)}{\d t}=a_\str(X(t))+b(X(t))\eta(t)
\label{str}
\eeq
can be essentially thought of as an ordinary differential equation.
It is amenable to non-linear changes of variable
according to the usual rules of differential and integral calculus.
The main disadvantage is that $X(t)$ and $\eta(t)$
at the same time $t$ are not independent.

\item {\it It\^o prescription.}
The Langevin-It\^o differential equation
\beq
\iito\frac{\d X(t)}{\d t}=a_\ito(X(t))+b(X(t))\eta(t)
\label{ito}
\eeq
has the advantage that the process $X(t)$
and the noise $\eta(t)$ at the same time $t$ are independent,
so that one has e.g.~$\mean{a_\ito(X)}=0$ in the stationary state.
Equation~(\ref{ito}) also provides a natural discretization of the process $X(t)$.
Considering discrete times $t=n\eps$ so that $X_n\equiv X(t_n)$,
we obtain the recursion
\beq
X_{n+1}=X_n+a_\ito(X_n)\eps+b(X_n)\zeta_{n+1},
\label{xrec}
\eeq
where $\zeta_{n+1}$ is a Gaussian random variable, independent of $X_n$,
such that $\mean{\zeta_{n+1}}=0$ and $\mean{\zeta_{n+1}^2}=\eps$.
This discrete scheme can be efficiently used in a numerical simulation.
The main disadvantage of the It\^o prescription
is that care must be exercised when making non-linear changes of variable.

\end{itemize}

Both Langevin equations~(\ref{str}) and~(\ref{ito}) describe
the same stochastic process $\{X(t)\}$
if their drift terms are related to each other by the correspondence formula
\beq
a_\ito(X)-a_\str(X)=\frac{1}{4}\,\frac{\d}{\d X}\,b(X)^2
=\frac{1}{2}\,b(X)\frac{\d b(X)}{\d X}.
\label{aias}
\eeq
The corresponding time-dependent probability density $P(x,t)=\mean{\delta(X(t)-x)}$
obeys the Fokker-Planck equation
\beqa
\frac{\dpar P}{\dpar t}
&=&-\diff{x}\Bigl(a_\ito(x)P\Bigr)
+\frac{1}{2}\,\frac{\dpar^2}{\dpar x^2}\Bigl(b(x)^2P\Bigr)\nonumber\\
&=&-\diff{x}\Bigl(a_\str(x)P\Bigr)
+\frac{1}{2}\,\diff{x}\Bigl(b(x)\diff{x}\Bigl(b(x)P\Bigr)\Bigr).
\eeqa
The stationary density of the process, $f_X(x)\equiv\lim_{t\to\infty}P(x,t)$, reads
\beq
f_X(x)
=\frac{N_\ito}{b(x)^2}\exp\left(2\int_{x_0}^x\frac{a_\ito(y)}{b(y)^2}\,\d y\right)
=\frac{N_\str}{b(x)}\exp\left(2\int_{x_0}^x\frac{a_\str(y)}{b(y)^2}\,\d y\right),
\label{pstat}
\eeq
where $x_0$ is an arbitrary initial point,
and $N_\ito$ and $N_\str$ are normalization constants.

It is now time to return to our functional $X$ (see~(\ref{xdef})).
The Langevin-Stratonovich equation~(\ref{xintro}), i.e.,
\beq
\sstr\frac{\d X(t)}{\d t}=1-X(t)+gX(t)\eta(t),
\label{xstr}
\eeq
is equivalent to the Langevin-It\^o equation
\beq
\iito\frac{\d X(t)}{\d t}=1+(D-1)X(t)+gX(t)\eta(t).
\label{xito}
\eeq
Indeed $b(X)=gX$ and $a_\str(X)=1-X$ yield $a_\ito(X)=1+(D-1)X$ (see~(\ref{ddef})).

The Langevin-Stratonovich equation~(\ref{xstr}) can be readily integrated.
We thus obtain the explicit stochastic representation of the process~as
\beq
X(t)=X(0)\,\e^{-t+gB(t)}+\int_0^t\e^{-(t-s)+g(B(t)-B(s))}\,\d s.
\label{xt}
\eeq
In the $t\to\infty$ limit,
the above expression loses the memory of its initial condition $X(0)$.
Moreover, using the stationarity of Brownian motion,
the integral can be recast as an integral over $\tau=t-s$,
which identifies with~(\ref{xdef2}) or~(\ref{xdef}).
We have thus shown that the exponential functional $X$
represents the stationary solution of~(\ref{xstr}) or~(\ref{xito}),
i.e., $X\equiv\lim_{t\to\infty}X(t)$.

The It\^o prescription allows us to directly read off the stationary mean value of $X$
from~(\ref{xito}):
\beq
\mean{X}=\frac{1}{1-D}.
\label{xmean}
\eeq
This expression only makes sense for $D<1$.
It is in agreement with the behavior of the time-dependent mean value $\mean{X(t)}$.
Equation~(\ref{xito}) indeed yields
\beq
\frac{\d\mean{X(t)}}{\d t}=1+(D-1)\mean{X(t)},
\eeq
whose solution, i.e.,
\beq
\mean{X(t)}=\frac{1}{1-D}+\left(X(0)-\frac{1}{1-D}\right)\e^{(D-1)t},
\label{meanxt}
\eeq
relaxes exponentially fast to~(\ref{xmean}) for $D<1$,
whereas it diverges for $D>1$.
The result~(\ref{meanxt}) can be recovered by averaging~(\ref{xt})
over the Brownian motion $\{B(s)\}$.

Equation~(\ref{pstat}) leads to the following result
for the distribution of the functional $X$:
\beq
f_X(x)=\frac{1}{D^{1/D}\Gamma(1/D)}\,x^{-(1+1/D)}\,\e^{-1/(Dx)}.
\label{fxres}
\eeq
This expression~\cite{yor,bouchaud,opper}
falls off exponentially fast as $x\to0$.
It exhibits a fat tail at large~$x$,
as it decreases as a power law, with the continuously variable exponent
$-(1+1/D)$.
As a consequence, the moment $\mean{X^s}$
only converges for $s<1/D$~(see~(\ref{mres})).
This provides another way to explain the divergence of~(\ref{xmean}) at $D=1$.

Finally, at par with the qualitative discussion given in the Introduction,
one can interpret the emergence of a fat tail at large values of $X$
and the possible divergence of the mean value $\mean{X}$
as consequences of the exchange of stability between the two fixed points
of the dynamical system~(\ref{xstr}),
i.e., the unstable deterministic one at $X=1$
and a fluctuating one sitting formally at $X=\infty$.

\begin{figure}[!ht]
\begin{center}
\includegraphics[angle=-90,width=.45\linewidth]{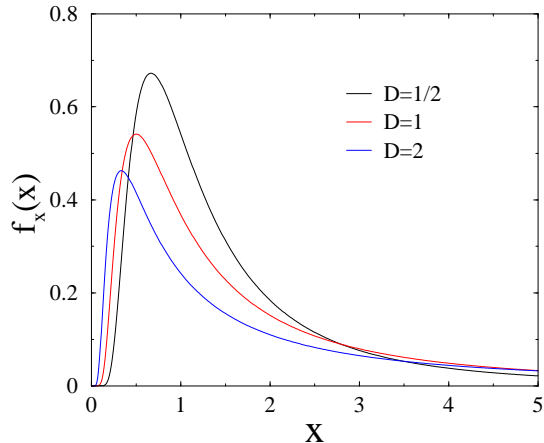}
\caption{\label{fx}
Plot of the probability density $f_X$ of the real functional $X$
(see~(\ref{fxres})), for three values of the diffusion constant $D$.}
\end{center}
\end{figure}

Figure~\ref{fx} shows a plot of the density~$f_X$ for three values of $D$.

\subsection{Kesten approach}
\label{realkes}

An alternative approach to study the random variable $X$
consists in using the similarity between the random integral~(\ref{xdef2})
and random sums of products referred to as Kesten variables.
Let us start with a reminder.
A Kesten variable~\cite{kesten,vervaat,dh,calan,theo} is defined~as
\beq
Z=1+\xi_1+\xi_1\xi_2+\xi_1\xi_2\xi_3+\cdots,
\label{kdef}
\eeq
where the $\xi_n$ are i.i.d.~positive random variables
with probability density $f_\xi(\xi)$.
If the latter distribution is such that $\mean{\ln\xi}<0$,
the sum~(\ref{kdef}) is almost surely convergent,
and it represents the stationary solution of the random recursion
\beq
Z_{n+1}=1+\xi_{n+1}Z_n,
\label{krec}
\eeq
where $\xi_{n+1}$ is independent of $Z_n$ and has density $f_\xi(\xi)$.
In other words, we have the identity among random variables
\beq
Z\equiv 1+\xi Z',
\label{ziden}
\eeq
where $Z'$ is a copy of $Z$, and $\xi$ is independent of $Z'$.
The density $f_Z(z)$ of the Kesten variable $Z$
obeys the integral equation
\beq
f_Z(z)=\int_0^\infty\frac{\d\xi}{\xi}\,f_\xi(\xi)\,
f_Z\!\left(\frac{z-1}{\xi}\right),
\label{dyson}
\eeq
which cannot be solved in closed form in general.

Let $\xi_\min=a$ and $\xi_\max=b$ be the smallest and largest values of $\xi$
(lower and upper bounds of the support of $f_\xi$),
and similarly $Z_\min=A$ and $Z_\max=B$ the smallest and largest values of $Z$.
The condition $\mean{\ln\xi}<0$ implies $a<1$.
We have then $A=1/(1-a)$.
If $a<b<1$, we have $B=1/(1-b)$ and $f_Z$ has finite support.
In the more interesting situation where $a<1<b$, the support of $f_Z$ extends to infinity.
It is known that the distribution $f_Z$ generically exhibits a fat tail,
i.e., a power-law fall-off, of the form
\beq
f_Z(z)\sim z^{-(1+\alpha)},
\eeq
where the exponent $\alpha>0$ is given by the condition~\cite{kesten,dh,calan}
\beq
\mean{\xi^\alpha}=1.
\label{xicond}
\eeq
The density $f_Z(z)$ has been derived explicitly~\cite{vervaat,calan}
in the case where $f_\xi(\xi)$
is a power law on an interval of the form $[0,b]$ or $[a,\infty]$.

The exponential functional $X$,
defined in~(\ref{xdef2}) as an integral of an exponential,
can be viewed as a continuous analogue of the Kesten variable~$Z$,
defined in~(\ref{kdef}) as a sum of products.
Similarly, the Langevin equation~(\ref{xstr}) is the continuous time analogue
of the recursion~(\ref{krec}).
This correspondence, already referred to in~\cite{comtet,bouchaud},
can be understood quantitatively in the following way.
Let us introduce a small time step~$\eps$ and split the integral
in~(\ref{xdef2}) as $X=X\sup{1}+X\sup{2}$, where
\beq
X\sup{1}=\int_0^\eps\e^{-t+gB(t)}\d t,\qquad
X\sup{2}=\int_\eps^\infty\e^{-t+gB(t)}\d t.
\eeq
To first order in $\eps$, we have $X\sup{1}=\eps$.
Furthermore, setting $t=\eps+s$, we have $B(t)\equiv B(\eps)+B(s)$,
so that $X\sup{2}=\e^{-\eps+\zeta}X'$,
where $X'$ is a copy of the variable $X$
and $\zeta=gB(\eps)$ is a Gaussian variable,
independent of $X'$, such that $\mean{\zeta}=0$
and $\mean{\zeta^2}=g^2\eps=2D\eps$.
Putting everything together, the exponential functional $X$ appears
(in the $\eps\to0$ limit) to obey the identity
\beq
X\equiv\eps+\xi X',
\label{xiden}
\eeq
with
\beq
\xi=\e^{-\eps+\zeta},
\label{xidef}
\eeq
where $\zeta$ is Gaussian, such that $\mean{\zeta}=0$ and
$\mean{\zeta^2}=2D\eps$.
Up to an unimportant global factor $\eps$,
the exponential functional $X$ therefore identifies
(in the $\eps\to0$ limit) with the Kesten variable $Z$
generated by the random input variables $\xi$ given by~(\ref{xidef}).
This is precisely the Kesten variable investigated in~\cite{theo},
where the main emphasis is already put on the $\eps\to0$ limit,
and where the distribution~(\ref{fxres}) is derived in this limit.

The above correspondence directly yields the fall-off exponent of the density of $X$.
We have indeed $\mean{\e^{s\zeta}}=\e^{s^2\mean{\zeta^2}/2}=\e^{Ds^2\eps}$, so that
\beq
\mean{\xi^s}=\e^{s(Ds-1)\eps}.
\label{xis}
\eeq
The condition~(\ref{xicond}) thus predicts $\alpha=1/D$, in agreement
with~(\ref{fxres}).

The full distribution of $X$ can actually be derived from
the identity~(\ref{xiden}).
Consider indeed the moment function
\beq
M(s)=\mean{X^s}.
\eeq
Equations~(\ref{xiden}) and~(\ref{xis}) can be respectively expanded
in the $\eps\to0$ limit to yield
\beq
M(s)=\mean{\xi^s}M(s)+s\mean{\xi^{s-1}}M(s-1)\eps+\cdots
\label{mexpand}
\eeq
and
\beq
\mean{\xi^s}=1+s(Ds-1)\eps+\cdots,
\eeq
where the dots stand for terms of higher order in $\eps$.
We are left (in the $\eps\to0$ limit) with the functional equation
\beq
M(s-1)=(1-Ds)M(s),
\label{fun}
\eeq
whose normalized solution reads
\beq
M(s)=\frac{\Gamma(-s+1/D)}{\Gamma(1/D)D^s}\qquad(\Re s<1/D).
\label{mres}
\eeq
Setting $s=1$,
we recover the expression~(\ref{xmean}) for $\mean{X}$, provided $D<1$.
The density of $X$ is given by the inverse Mellin transform
\beq
f_X(x)=\int\frac{\d s}{2\pi\i}\,x^{-s-1} M(s).
\eeq
The expression~(\ref{fxres}) is recovered by summing the contributions of the
poles of the integrand at $s=1/D+n$ for $n=0,1,\dots$
The leftmost pole at $s=1/D$ is responsible for the power-law tail
with exponent $-(1+1/D)$.

\section{The imaginary functional}
\label{imag}

We now turn to the main object of this paper,
namely the distribution of the random variable $Z$
defined by the integral~(\ref{zdef}), i.e.,
\beq
Z=\int_0^\infty\e^{-t+\i gB(t)}\d t.
\label{zdef2}
\eeq
From a formal viewpoint,
the complex variable $Z$ can be viewed as the analytical continuation
as $g\to\i g$ of its real counterpart $X$, investigated in Section~\ref{real}.
This continuation amounts to changing
the sign of the diffusion constant ($D\to-D$).
It is however worth emphasizing that the study of $Z$
is far more difficult than that of~$X$,
as most of the usual tools which are fit to investigate real random variables
cease to work in the case of a complex random variable.

Setting $Z=X+\i Y$, we are equivalently interested in the joint
distribution $f(x,y)$ of the two correlated real random variables
\beq
X=\int_0^\infty\e^{-t}\,\cos(gB(t))\,\d t,\qquad
Y=\int_0^\infty\e^{-t}\,\sin(gB(t))\,\d t.
\label{xydef}
\eeq
Let us start with a few general facts.
The complex random variable $Z$ lives inside the unit disk, since
\beq
\abs{Z}\le\int_0^\infty\e^{-t}\,\d t=1.
\eeq
Let us anticipate that its distribution has a smooth density $f(x,y)$,
whose support is the whole unit disk.
Before we turn to more specific features,
it is illustrative to first evaluate the mean~$\mean{Z}$.
Using the identity
\beq
\mean{\e^{\i gB(t)}}=\e^{-(g^2/2)\mean{B(t)^2}}=\e^{-Dt},
\eeq
we obtain at once
\beq
\mean{Z}=\int_0^\infty\e^{-t}\mean{\e^{\i gB(t)}}\d t
=\int_0^\infty\e^{-(D+1)t}\,\d t,
\eeq
i.e.,
\beq
\mean{Z}=\frac{1}{D+1},
\label{zmean}
\eeq
so that
\beq
\mean{X}=\frac{1}{D+1},\qquad\mean{Y}=0.
\label{xymean}
\eeq
The above results obey the conjugation symmetry:
$Z=X+\i Y$ and $\Z=X-\i Y$ have the same law.
The result~(\ref{zmean}) can be recovered
by performing the analytical continuation $D\to-D$ on~(\ref{xmean}).
At variance with the latter result, the expression~(\ref{zmean})
depends smoothly on the diffusion constant $D$,
decreasing from $\mean{Z}=1$ in the $D\to0$ limit
to $\mean{Z}\to0$ in the $D\to\infty$ limit.
These limiting regimes will be respectively investigated
in Sections~\ref{weak} and~\ref{strong},
whereas the main features of the dependence of the distribution of $Z$
on $D$ will be studied in Section~\ref{dep}.

\subsection{Langevin approach}
\label{lan}

It can be shown along the lines of Section~\ref{reallan}
that the random variable $Z\equiv\lim_{t\to\infty}Z(t)$
represents the stationary solution of the Langevin-Stratonovich equation
\beq
\sstr\frac{\d Z}{\d t}=1-Z+\i gZ\eta(t),
\label{zstr}
\eeq
where $\eta(t)$ is again a Gaussian white noise.
The solution to~(\ref{zstr}) reads
\beq
Z(t)=Z(0)\,\e^{-t+\i gB(t)}+\int_0^t\e^{-(t-s)+\i g(B(t)-B(s))}\,\d s.
\label{zt}
\eeq
A comparison with~(\ref{Phisol}) shows that the identity~(\ref{PhiZ})
extends to finite times as $\zed(t)\equiv 2Z(t/2)$ (again with $D=1$).

The Langevin-It\^o equation corresponding to~(\ref{zstr}) is
\beq
\iito\frac{\d Z}{\d t}=1-(D+1)Z+\i gZ\eta(t).
\label{zito}
\eeq
This equation yields $\mean{1-(D+1)Z}=0$ in the stationary state.
We thus readily recover the expression~(\ref{zmean}) of $\mean{Z}$.
Equation~(\ref{zito}) also provides an efficient discrete scheme.
Considering discrete times $t=n\eps$,
and adapting the recursion~(\ref{xrec}) to the present situation, we get
\beq
Z_{n+1}=\eps+\hat\xi_{n+1}Z_n,
\label{zrec1}
\eeq
with
\beq
\hat\xi_n=1-(D+1)\eps+\i\zeta_n,
\eeq
and where $\zeta_n$ is a Gaussian random variable,
such that $\mean{\zeta_n}=0$ and $\mean{\zeta_n^2}=2D\eps$.

\begin{figure}[!ht]
\begin{center}
\includegraphics[angle=-90,width=.45\linewidth]{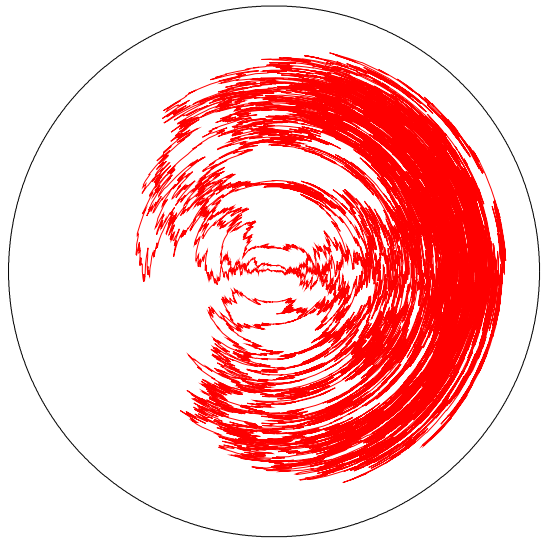}
\caption{\label{tra}
A typical trajectory $\{Z(t)\}$ of length $t=25$ in the unit disk,
for $D=1$, i.e., the case of the binary annihilation process~(\ref{RDS}).}
\end{center}
\end{figure}

Figure~\ref{tra} shows a typical trajectory $\{Z(t)\}$
starting from the origin, of length $t=25$,
for $D=1$, i.e., the case of the binary annihilation process~(\ref{RDS}).
The trajectory is generated by means of the scheme~(\ref{zrec1}),
with $\eps=10^{-3}$.
The main advantage of using the complex coordinate $Z(t)$
(instead of the original field variable $\varphi(t)$~\cite{mun,delou})
is that unbounded excursions are avoided, as we have $\abs{Z(t)}\le1$.
The rest of this section is devoted to the distribution of the functional $Z$,
i.e., equivalently, of the generic point of such a trajectory in the long-time regime.

\subsection{Kesten approach}
\label{kes}

The random variable $Z$ can be alternatively described as a complex Kesten variable.
Along the lines of Section~\ref{realkes},
it can indeed be shown that $Z$ obeys the identity
(in the $\eps\to0$ limit)
\beq
Z\equiv\eps+\xi Z',
\label{Ziden}
\eeq
where $Z'$ is a copy of $Z$, whereas
\beq
\xi=\e^{-\eps+\i\zeta},
\eeq
and $\zeta$ is Gaussian,
such that $\mean{\zeta}=0$ and $\mean{\zeta^2}=2D\eps$.
In other words, $Z$ represents (in the $\eps\to0$ limit)
the stationary solution of the random recursion
\beq
Z_{n+1}=\eps+\xi_{n+1}Z_n,
\label{zrec2}
\eeq
with
\beq
\xi_n=\e^{-\eps+\i\zeta_n}.
\eeq
The variable $Z$ therefore identifies
(up to an unimportant global factor $\eps$, and in the $\eps\to0$ limit)
with the Kesten variable generated by the complex input variables $\xi_n$.

It is worth noticing the close resemblance between
the random recursions~(\ref{zrec1}) and~(\ref{zrec2}),
respectively corresponding to the Langevin and Kesten approaches.
Both discrete schemes become equivalent to leading order
in the $\eps\to0$ limit.
The Langevin scheme involving $\hat\xi_n$ appears in this regime
as a suitably linearized form of the non-linear Kesten scheme involving $\xi_n$.
The Langevin scheme is more suitable for extensive numerical simulations,
as it does not involve the complex exponential function.

\subsection{Moments}
\label{mom}

In the case of the real exponential functional $X$,
the explicit form~(\ref{fxres}) of the distribution $f_X$
could be derived either from the Langevin approach,
as the stationary solution~(\ref{pstat}) of the Fokker-Planck equation,
or from the Kesten approach, by means of the identity~(\ref{xiden})
in the $\eps\to0$ limit.
In the present case of the imaginary exponential functional $Z$,
neither route leads to an explicit expression for the density $f(x,y)$.

The Kesten approach however directly yields an interesting piece of information,
in the form of a recursion relation for the moments $\mean{Z^k\Z^l}$.
We recall that the distribution
of a real random variable $X$ is entirely characterized
by the single array of moments $\mean{X^k}$ ($k=0,1,\dots$)
(leaving aside questions related to convergence),
whereas a double array of moments $\mean{Z^k\Z^l}$
(or, equivalently, $\mean{X^kY^l}$) is needed
to characterize the distribution of a complex random variable $Z=X+\i Y$.

The analysis of the moments
follows the line of thought of Section~\ref{realkes}.
For $k=0,1,\dots$ integer, we expand the identity~(\ref{Ziden}) as
\beq
Z^k=\xi^k(Z')^k+k\xi^{k-1}(Z')^{k-1}\eps+\cdots
\eeq
We thus obtain an expansion similar to~(\ref{mexpand}), i.e.,
\beqa
\mean{Z^k\Z^l}&=&\mean{\xi^k\xxi^l}\mean{Z^k\Z^l}\nonumber\\
&+&\Bigl(k\mean{\xi^{k-1}\xxi^l}\mean{Z^{k-1}\Z^l}
+l\mean{\xi^k\xxi^{l-1}}\mean{Z^k\Z^{l-1}}\Bigr)\eps+\cdots
\label{momexpand}
\eeqa
Furthermore, we have
\beqa
\mean{\xi^k\xxi^l}=\e^{-(k+l)\eps}\mean{\e^{\i(k-l)\zeta}}
&=&\e^{-(k+l+(k-l)^2D)\eps}\nonumber\\
&=&1-(k+l+(k-l)^2D)\eps+\cdots
\eeqa
Inserting the latter expansion into~(\ref{momexpand}),
we are left (in the $\eps\to0$ limit) with the recursion
\beq
(k+l+(k-l)^2D)\mean{Z^k\Z^l}=k\mean{Z^{k-1}\Z^l}+l\mean{Z^k\Z^{l-1}}.
\label{momrec}
\eeq

The formula~(\ref{momrec}) is one of the key results of this work.
It allows one to evaluate the moments $\mean{Z^k\Z^l}$ in a recursive way.
All of them are rational functions of the diffusion constant $D$.
They are regular over the whole range of positive values of $D$.
Owing to the symmetry between $Z$ and $\Z$, we have
$\mean{Z^k\Z^l}=\mean{Z^l\Z^k}$,
so that we can restrict ourselves to $k\ge l$.
The expressions of the first moments
(for $k+l\le6$) are listed in Table~\ref{momtab}.

\begin{table}[!ht]
\begin{center}
\begin{tabular}{|c|c|}
\hline
moment&expression\\
\hline
$\mean{Z}$&$\frac{1}{D+1}$\\
\hline
$\mean{Z^2}$&$\frac{1}{(D+1)(2D+1)}$\\
$\mean{Z\Z}$&$\frac{1}{D+1}$\\
\hline
$\mean{Z^3}$&$\frac{1}{(D+1)(2D+1)(3D+1)}$\\
$\mean{Z^2\Z}$&$\frac{4D+3}{(D+1)(D+3)(2D+1)}$\\
\hline
$\mean{Z^4}$&$\frac{1}{(D+1)(2D+1)(3D+1)(4D+1)}$\\
$\mean{Z^3\Z}$&$\frac{9D^2+10D+3}{(D+1)^2(D+3)(2D+1)(3D+1)}$\\
$\mean{Z^2\Z^2}$&$\frac{4D+3}{(D+1)(D+3)(2D+1)}$\\
\hline
$\mean{Z^5}$&$\frac{1}{(D+1)(2D+1)(3D+1)(4D+1)(5D+1)}$\\
$\mean{Z^4\Z}$&$\frac{16D^2+13D+3}{(D+1)^2(D+3)(2D+1)(3D+1)(4D+1)}$\\
$\mean{Z^3\Z^2}$&$\frac{36D^3+93D^2+68D+15}{(D+1)^2(D+3)(D+5)(2D+1)(3D+1)}$\\
\hline
$\mean{Z^6}$&$\frac{1}{(D+1)(2D+1)(3D+1)(4D+1)(5D+1)(6D+1)}$\\
$\mean{Z^5\Z}$&$\frac{25D^2+16D+3}{(D+1)^2(D+3)(2D+1)(3D+1)(4D+1)(5D+1)}$\\
$\mean{Z^4\Z^2}$&$\frac{288D^3+544D^2+279D+45}{(D+1)(D+3)(D+5)(2D+1)(2D+3)(3D+1)(4D+1)}$\\
$\mean{Z^3\Z^3}$&$\frac{36D^3+93D^2+68D+15}{(D+1)^2(D+3)(D+5)(2D+1)(3D+1)}$\\
\hline
\end{tabular}
\end{center}
\caption{Expressions of the moments $\mean{Z^k\Z^l}$
for $k+l\le6$ as a function of the diffusion constant $D$.}
\label{momtab}
\end{table}

The moments $\mean{Z^k}$ can be evaluated in closed form.
Equation~(\ref{momrec}) indeed boils down for $l=0$ to the recursion
\beq
(kD+1)\mean{Z^k}=\mean{Z^{k-1}},
\eeq
which can be viewed as the analytical continuation of the functional equation~(\ref{fun}).
We thus obtain
\beq
\mean{Z^k}=\frad{1}{\prod_{j=1}^k(jD+1)}=\frac{\Gamma(1+1/D)}{D^k\,\Gamma(k+1+1/D)}.
\label{zmom}
\eeq
For $D=1$, the above expression simplifies to
\beq
\label{Zk}
\mean{Z^k}\big{|}_{D=1}=\frac{1}{(k+1)!}.
\eeq
This result also holds for negative $k$.
Setting $k=-p$, with $p\ge2$ integer, we obtain $\mean{Z^{-p}}\big{|}_{D=1}=0$.
We thus recover a result mentioned in the Introduction,
concerning the null factorial moments for the number of particles in the steady state
for the binary annihilation process~(\ref{RDS}).

The joint moments $\mean{X^kY^l}$ of the real variables $X$ and $Y$
can be obtained as linear combinations of the moments $\mean{Z^k\Z^l}$,
using $X=(Z+\Z)/2$ and $Y=\i(\Z-Z)/2$.
The conjugation symmetry ensures that $f(x,y)=f(x,-y)$,
so that $\mean{X^kY^l}=0$ whenever~$l$ is odd.
The expressions of the first non-vanishing moments $\mean{X^kY^l}$
(for $l$ even and $k+l\le6$) are listed in Table~\ref{mxytab}.

\begin{table}[!ht]
\begin{center}
\begin{tabular}{|c|c|}
\hline
moment&expression\\
\hline
$\mean{X}$&$\frac{1}{D+1}$\\
\hline
$\mean{X^2}$&$\frac{1}{2D+1}$\\
$\mean{Y^2}$&$\frac{D}{(D+1)(2D+1)}$\\
\hline
$\mean{X^3}$&$\frac{9D^2+10D+3}{(D+1)(D+3)(2D+1)(3D+1)}$\\
$\mean{XY^2}$&$\frac{3D}{(D+3)(2D+1)(3D+1)}$\\
\hline
$\mean{X^4}$&$\frac{18D^4+60D^3+58D^2+22D+3}{(D+1)^2(D+3)(2D+1)(3D+1)(4D+1)}$\\
$\mean{X^2Y^2}$&$\frac{D(6D^2+8D+3)}{(D+1)(D+3)(2D+1)(3D+1)(4D+1)}$\\
$\mean{Y^4}$&$\frac{3D^2(6D^2+8D+3)}{(D+1)^2(D+3)(2D+1)(3D+1)(4D+1)}$\\
\hline
$\mean{X^5}$&$\frac{450D^4+940D^3+606D^2+158D+15}{(D+1)(D+3)(D+5)(2D+1)(3D+1)(4D+1)(5D+1)}$\\
$\mean{X^3Y^2}$&$\frac{D(90D^4+268D^3+249D^2+98D+15)}{(D+1)^2(D+3)(D+5)(2D+1)(3D+1)(4D+1)(5D+1)}$\\
$\mean{XY^4}$&$\frac{3D^2(30D^3+86D^2+63D+15)}{(D+1)^2(D+3)(D+5)(2D+1)(3D+1)(4D+1)(5D+1)}$\\
\hline
$\mean{X^6}$&$\frac{2700D^7+16740D^6+35940D^5+36182D^4+19056D^3+5383D^2+774D+45}{(D+1)^2(D+3)(D+5)(2D+1)(2D+3)(3D+1)(4D+1)(5D+1)(6D+1)}$\\
$\mean{X^4Y^2}$&$\frac{3D(180D^5+756D^4+1044D^3+584D^2+148D+15)}{(D+1)(D+3)(D+5)(2D+1)(2D+3)(3D+1)(4D+1)(5D+1)(6D+1)}$\\
$\mean{X^2Y^4}$&$\frac{3D^2(180D^4+576D^3+638D^2+279D+45)}{(D+1)(D+3)(D+5)(2D+1)(2D+3)(3D+1)(4D+1)(5D+1)(6D+1)}$\\
$\mean{Y^6}$&$\frac{15D^3(180D^4+576D^3+638D^2+279D+45)}{(D+1)^2(D+3)(D+5)(2D+1)(2D+3)(3D+1)(4D+1)(5D+1)(6D+1)}$\\
\hline
\end{tabular}
\end{center}
\caption{Expressions of the moments $\mean{X^kY^l}$
for $l$ even and $k+l\le6$ as a function of the diffusion constant $D$.}
\label{mxytab}
\end{table}

\subsection{Fokker-Planck equation}
\label{fp}

The next natural step in our analysis consists in writing down
the Fokker-Planck equation obeyed by the distribution of the variable $Z$,
or, equivalently, by the joint distribution of the real variables $X$ and $Y$.
We shall successively do so using various representations.

We start by considering the two-variable characteristic function
\beq
\Phi(p,q)=\mean{\e^{pZ+q\Z}}=\sum_{k,l=0}^\infty\mean{Z^k\Z^l}\frac{p^kq^l}{k!l!}.
\eeq
The recursion~(\ref{momrec}) is equivalent to the partial differential equation
\beq
\left(D\left(p\diff{p}-q\diff{q}\right)^2+p\diff{p}+q\diff{q}-p-q\right)\Phi(p,q)=0,
\label{pqpde}
\eeq
with $\Phi(0,0)=1$.
For $q=0$, the above equation boils down to the solvable ordinary differential equation
\beq
Dp\phi''(p)+(D+1)\phi'(p)-\phi=0
\eeq
for the characteristic function $\phi(p)=\mean{\e^{pZ}}$,
where accents denote differentiations.
We thus obtain
\beq
\phi(p)
=\Gamma(1+1/D)\left(\frac{p}{D}\right)^{-1/(2D)}I_{1/D}\left(2\sqrt\frac{p}{D}\right),
\eeq
where $I_{1/D}$ is the modified Bessel function.
This result agrees with the expression~(\ref{zmom}) of the moments
$\mean{Z^k}$.

The characteristic function $\Psi(u,v)=\mean{\e^{uX+vY}}$
is equal to $\Phi(p,q)$, with $p=(u-\i v)/2$ and $q=(u+\i v)/2$.
It therefore obeys the partial differential equation
\beq
\left(D\left(u\diff{v}-v\diff{u}\right)^2-u\diff{u}-v\diff{v}+u\right)\Psi(u,v)=0,
\label{uvpde}
\eeq
with $\Psi(0,0)=1$.
The latter equation is equivalent to the following equation
for the joint density of $X,Y$:
\beq
\left(D\left(y\diff{x}-x\diff{y}\right)^2+(x-1)\diff{x}+y\diff{y}+2\right)f(x,y)=0.
\label{xypde}
\eeq
The above equation can be recovered by noticing that
$f(x,y)$ is the stationary solution
of the two-dimensional Fokker-Planck equation associated with the two coupled
Langevin-Stratonovich equations
\beqa
\frac{\d X(t)}{\d t}&=&1-X(t)-gY(t)\eta(t),\nonumber\\
\frac{\d Y(t)}{\d t}&=&-Y(t)+gX(t)\eta(t).
\label{reim}
\eeqa
A single real noise $\eta(t)$ enters both equations~(\ref{reim}),
and so the above system of Langevin equations is degenerate.
As a consequence, the second-order differential operators which appear in
either~(\ref{pqpde}),~(\ref{uvpde}), or~(\ref{xypde}) are of parabolic nature.
This phenomenon is best exhibited in yet another form,
using the polar coordinates $Z=R\e^{\i\theta}$.
The system~(\ref{reim}) then becomes
\beqa
\frac{\d R(t)}{\d t}&=&-R(t)+\cos\theta(t),\nonumber\\
\frac{\d\theta(t)}{\d t}&=&-\frac{\sin{\theta(t)}}{R(t)}+g\eta(t),
\label{rt}
\eeqa
showing explicitly that the noise only acts in the angular direction.
This property was already noticed in~\cite{mun,delou}
in the context of binary annihilation~(see~(\ref{RDS})),
i.e., for $D=1$.

Coming back to the stationary state,
still denoting $Z=R\e^{\i\theta}$, and setting
\beq
p=\frac{\mu}{2}\,\e^{-\i\nu},\qquad
q=\frac{\mu}{2}\,\e^{\i\nu},
\eeq
the characteristic function $\Phi(p,q)$ becomes the generating function
\beq
\Omega(\mu,\nu)=\mean{\e^{\mu X_\nu}},
\label{omegadef}
\eeq
where
\beq
X_\nu=\frac{1}{2}(Z\e^{-\i\nu}+\Z\e^{\i\nu})
=R\cos(\theta-\nu)=X\cos\nu+Y\sin\nu
\label{xnudef}
\eeq
is the projection of $Z$ onto an axis at an angle $\nu$.
The function $\Omega(\mu,\nu)$ obeys
\beq
\left(D\frac{\dpar^2}{\dpar\nu^2}-\mu\diff{\mu}+\mu\cos\nu\right)\Omega(\mu,\nu)=0,
\label{munupde}
\eeq
with $\Omega(0,\nu)=1$ for all $\nu$.

The partial differential equation~(\ref{munupde})
is another central result of this work.
It will be used in the following
in order to investigate the density of $Z$ in various limits.
We shall discuss the weak-disorder regime ($D\to0$) in section~\ref{weak},
the strong-disorder limit ($D\to\infty$) in sections~\ref{strong} and~\ref{larged},
and the asymptotic behavior of the density of~$Z$ near the unit circle
in section~\ref{circle}.
In spite of its relatively simple form,
we have not been able to solve the parabolic
partial differential equation~(\ref{munupde}) in full generality.
We nevertheless find it worth to sketch below two approaches we have tried.

\subsubsection*{Fourier-series expansion.}

The generating function $\Omega(\mu,\nu)$ is $2\pi$-periodic in the angular variable $\nu$.
It is henceforth natural to expand it as a Fourier series:
\beq
\label{Fourier}
\Omega(\mu,\nu)=\sum_{n=-\infty}^\infty\omega_n(\mu)\,\e^{\i n\nu}.
\eeq
The Fourier coefficients have the symmetry
$\omega_{-n}(\mu)=\bar{\omega}_n(\mu)$.
They obey the coupled differential equations
\beq
\label{difdif}
\left(Dn^2+\mu\frac{\d}{\d\mu}\right)\omega_n(\mu)=
\frac{\mu}{2}\left(\omega_{n+1}(\mu)+\omega_{n-1}(\mu)\right),
\eeq
with $\omega_n(0)=\delta_{n,0}$.
For $D=0$,~(\ref{difdif}) coincides with the differential recursion relation
obeyed by the modified Bessel functions $I_n(\mu)$.
Looking for asymptotics similar to those of Bessel functions,
we obtain the following behavior of the amplitudes $\omega_n(\mu)$
at small and large values of $\mu$, for all $n\ge0$:
\beqa
\omega_n(\mu)&\approx&\frac{\Gamma(1+1/D)}{n!\,\Gamma(n+1+1/D)}
\left(\frac{\mu}{2D}\right)^n\qquad(\mu\ll1),\nonumber\\
\omega_n(\mu)&\sim&\e^{\mu}{\hskip 153pt}(\mu\gg1).
\eeqa
A more advanced analysis of the large-$\mu$ regime will be performed
in Section~\ref{circle}.

\subsubsection*{Expansion in trigonometric polynomials.}

The generating function $\Omega(\mu,\nu)$
can alternatively be expanded as a power series in $\mu$ of the form
\beq
\Omega(\mu,\nu)=\sum_{k=0}^{\infty}\frac{\mu^k}{k!}\,P_k(\cos\nu).
\eeq
Equation~(\ref{omegadef}) yields
\beq
P_k(\cos\nu)=\mean{X_\nu^k}=\mean{(X\cos\nu+Y\sin\nu)^k}
=\frac{1}{2^k}\mean{(Z\e^{-\i\nu}+\Z\e^{\i\nu})^k}.
\label{pkdef}
\eeq
The $P_k$ are polynomials of degree $k$ in $x=\cos\nu$.
They satisfy the differential recursion relation
\beq
\label{recurP}
D((x^2-1)P''_k+xP'_k)+kP_k=kx P_{k-1},
\eeq
with $P_0(x)=1$.
Alternatively, as a consequence of~(\ref{pkdef}), we have
\beq
P_k(\cos\nu)=\frac{1}{2^k}\sum_{j=0}^k{k\choose j}
\mean{Z^{k-j}\Z^j}\cos((2j-k)\nu).
\label{pkexpand}
\eeq
Both~(\ref{recurP}) and~(\ref{pkexpand}) together with~(\ref{momrec})
or Table~\ref{momtab} consistently yield
\beqa
P_1(x)=\frac{x}{D+1},\qquad
P_2(x)=\frac{x^2+D}{(D+1)(2D+1)},\nonumber\\
P_3(x)=\frac{(D+3)x^3+9D(D+1)x}{(D+1)(D+3)(2D+1)(3D+1)},
\eeqa
and so on.

The polynomials $P_k(x)$ do not seem to belong to any
of the known standard families of polynomials.
In particular, they do not form a set of orthogonal polynomials.
This can easily be checked using the fact
that orthogonal polynomials must obey a three-term linear recursion~\cite{szego}.

An asymptotic analysis of the large-order behavior of the polynomials $P_k(x)$
will also be done in Section~\ref{circle}.

\subsection{Weak-disorder regime ($D\to0$)}
\label{weak}

In this regime,
the complex exponential involved in~(\ref{zdef}) or~(\ref{zdef2})
or, equivalently, the trigonometric functions involved in~(\ref{xydef}),
can be expanded in powers of $g\equiv\sqrt{2D}$.
We thus obtain
\beq
Z=1+\i\sqrt{D}\,U-DV+\cdots,
\eeq
i.e.,
\beq
X=1-DV+\cdots,\qquad
Y=\sqrt{D}\,U+\cdots,
\label{scadef}
\eeq
where $U$ and $V$ are the following random integrals:
\beq
U=\sqrt{2}\int_0^\infty\e^{-t}\,B(t)\,\d t,\qquad
V=\int_0^\infty\e^{-t}\,B(t)^2\,\d t.
\label{uvdef}
\eeq
The variable $U$ is Gaussian,
and therefore entirely characterized by its second moment $\mean{U^2}=1$.
This is however not the end of the story.
The variables $U$ and~$V$ indeed have a non-trivial joint distribution
supported by the parabolic domain $V\ge U^2/2$.
As these variables are respectively defined
as a linear and a quadratic functional of Brownian motion,
one way to investigate their joint distribution
consists in evaluating the corresponding characteristic function
as a Gaussian functional integral.
We prefer to use a more direct approach,
based on the solution of the partial differential equation~(\ref{munupde})
in the relevant regime, i.e., $D\ll1$ and $\nu\ll1$.
We are thus led to consider a simplified version of~(\ref{munupde}),
where the cosine function has been expanded to second order:
\beq
\left(D\frac{\dpar^2}{\dpar\nu^2}-\mu\diff{\mu}+\mu\left(1-\frac{\nu^2}{2}\right)\right)
\Omega(\mu,\nu)=0.
\label{simplepde}
\eeq
The latter equation can be solved for arbitrary $D$ if,
inspired by the property that $U$ is Gaussian,
we make a Gaussian Ansatz of the form
\beq
\Omega(\mu,\nu)=a(\mu)\e^{-b(\mu)\nu^2}.
\label{simple}
\eeq
We are left with two ordinary differential equations for $a(\mu)$ and $b(\mu)$, namely
\beq
b'(\mu)=\frac{1}{2}-\frac{4D}{\mu}\,b(\mu)^2,\qquad
a'(\mu)=\left(1-\frac{2D}{\mu}\,b(\mu)\right)a(\mu),
\eeq
with $a(0)=1$ and $b(0)=0$.
The equation for $b(\mu)$ is a Riccati equation, and is therefore solvable.
Indeed, setting
\beq
b(\mu)=-\frac{1}{8D}+\frac{\mu}{4D}\,\frac{\psi'(\mu)}{\psi(\mu)},
\eeq
we obtain for $\psi(\mu)$ the second-order linear differential equation
\beq
\psi''(\mu)=\left(\frac{2D}{\mu}-\frac{1}{4\mu^2}\right)\psi(\mu),
\eeq
whose regular solution is
\beq
\psi(\mu)=\sqrt{\mu}\,I_0\left(2\sqrt{2D\mu}\right).
\eeq
Skipping details, we obtain after some algebra
\beq
a(\mu)=\frac{\e^{\mu}}{\sqrt{I_0(2\sqrt{2D\mu})}},\qquad
b(\mu)=\frac{1}{4D}\sqrt{2D\mu}\,\frac{I_1(2\sqrt{2D\mu})}{I_0(2\sqrt{2D\mu})},
\label{absol}
\eeq
where $I_0$ and $I_1$ are modified Bessel functions.
In the following we use various properties of Bessel functions
(differential equations, series and asymptotic expansions, location of the zeros)
exposed e.g.~in~\cite{gr}.

Now, inserting~(\ref{scadef}) into~(\ref{omegadef}) and~(\ref{xnudef}),
and expanding the result for $D$ and $\nu$ small, we obtain
\beq
\Omega(\mu,\nu)
=\bigmean{\exp\left(\mu+\sqrt{D}\mu\nu U-D\mu
V-\frac{1}{2}\mu\nu^2+\cdots\right)}.
\eeq
In the limit where $\mu$ is large, at fixed values
of the combinations $\rho=-\sqrt{D}\mu\nu$ and $\sigma=D\mu$,
the leading term $\mu$ of the above expansion is divergent,
whereas all the other terms remain finite,
and the higher-order terms which are not written down vanish.
As a consequence, a direct identification with~(\ref{simple}),~(\ref{absol})
yields an explicit expression for the characteristic function
\beq
F(\rho,\sigma)
=\mean{\e^{-\rho U-\sigma V}}
=\frac{1}{\sqrt{I_0(2\sqrt{2\sigma})}}
\exp\left(\frac{I_2(2\sqrt{2\sigma})}{2\sigma
I_0(2\sqrt{2\sigma})}\,\rho^2\right).
\label{fres}
\eeq
This expression is another key result of this work.
It contains the full joint distribution of the rescaled variables $U$ and $V$.
Table~\ref{muvtab} gives the rational values of the first few
moments~$\mean{U^kV^l}$
(for $k$ even and $k,l\le4$).
The positive variables $U^2$ and $V$ appear as strongly positively correlated.

\begin{table}[!ht]
\begin{center}
\begin{tabular}{|c|c||c|c||c|c|}
\hline
moment&value&moment&value&moment&value\\
\hline
&&$\mean{U^2}$&$1$&$\mean{U^4}$&$3$\\
$\mean{V}$&$1$&$\mean{U^2V}$&$\frac{7}{3}$&$\mean{U^4V}$&$11$\\
$\mean{V^2}$&$2$&$\mean{U^2V^2}$&$\frac{25}{3}$&$\mean{U^4V^2}$&$\frac{164}{3}$\\
$\mean{V^3}$&$\frac{20}{3}$&$\mean{U^2V^3}$&$\frac{613}{15}$&$\mean{U^4V^3}$&$\frac{1726}{5}$\\
$\mean{V^4}$&$\frac{95}{3}$&$\mean{U^2V^4}$&$\frac{2305}{9}$&$\mean{U^4V^4}$&$\frac{39751}{15}$\\
\hline
\end{tabular}
\end{center}
\caption{Values of the moments $\mean{U^kV^l}$ for $k$ even and $k,l\le4$.}
\label{muvtab}
\end{table}

\begin{figure}[!ht]
\begin{center}
\includegraphics[angle=-90,width=.45\linewidth]{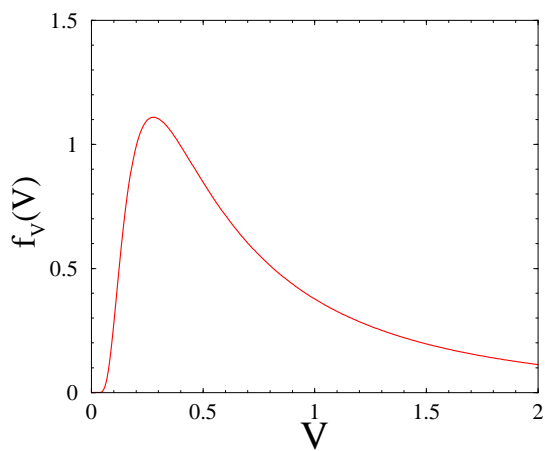}
\caption{\label{fv}
Plot of the probability density $f_V$ of the scaling variable $V$.}
\end{center}
\end{figure}

The Gaussian marginal distribution of $U$ is recovered
as $F(\rho,0)=\mean{\e^{-\rho U}}=\e^{\rho^2/2}$.
The marginal distribution of the quadratic functional $V$
is given by the inverse Laplace transform of $F(0,\sigma)$, i.e.,
\beq
f_V(v)=\int\frac{\d\sigma}{2\pi\i}\,\frac{\e^{\sigma
v}}{\sqrt{I_0(2\sqrt{2\sigma})}}.
\label{fvres}
\eeq
The behavior of this density at small $v$
is given by a saddle point at a large positive value
$\sigma\approx1/(2v^2)$ of the integration variable.
We thus obtain the exponential fall-off
\beq
f_V(v)\sim\e^{-1/(2v)}\qquad(v\to0).
\label{fvinf}
\eeq
Its behavior at large $v$
is determined by the rightmost singularity of the integrand in~(\ref{fvres}),
which is a branch-cut singularity at $\sigma=-\sigma_1$,
with $\sigma_1=j_1^2/8\approx0.722897$,
where $j_1\approx2.404825$ is the first zero of the Bessel function $J_0$.
We thus obtain the exponential fall-off
\beq
f_V(v)\sim\e^{-\sigma_1v}\qquad(v\to\infty).
\eeq
Figure~\ref{fv} shows a plot of the density $f_V(v)$,
obtained by means of a numerical integration of the formula~(\ref{fvres})
along the imaginary $\sigma$ axis.

\subsection{Strong-disorder regime ($D\to\infty$)}
\label{strong}

In this regime, the integrand in the expression~(\ref{zdef}) or~(\ref{zdef2})
involves a large and hence rapidly varying phase.
The variable $Z$ can therefore be expected to behave
as a complex Gaussian variable with zero mean and a small width.
Still denoting $Z=R\e^{\i\theta}$,
we have indeed $\mean{R^2}=\mean{Z\Z}\approx1/D$ (see~Table~\ref{momtab}).

The convergence of $Z$ to a complex Gaussian variable
can be studied by means of the moments investigated in Section~\ref{mom}.
The form of the recursion~(\ref{momrec}) implies that the moments behave as
$\mean{Z^k\Z^l}\sim1/D^{\max(k,l)}$ at large $D$.
As a consequence,
the diagonal moments $\mean{R^{2k}}=\mean{(Z\Z)^k}\sim1/D^k$
are the leading ones at large $D$,
whereas the non-diagonal ones
$\mean{Z^n(Z\Z)^k}\sim1/D^{k+n}$ ($n=1,2,\dots$) are more and more subleading.
More precisely, setting
\beq
\mean{Z^n(Z\Z)^k}=\frac{a\sup{n}_k}{D^{k+n}},
\eeq
the recursion~(\ref{momrec}) is equivalent to
\beqa
a\sup{0}_k=a\sup{1}_{k-1},\nonumber\\
(Dn^2+2k+n)a\sup{n}_k&=&D(k+n)a\sup{n-1}_k+ka\sup{n+1}_{k-1}\qquad(n\ge1).
\label{leadrec}
\eeqa
To leading order as $D\to\infty$, the latter equation boils down to
\beq
n^2a\sup{n}_k=(k+n)a\sup{n-1}_k,
\eeq
so that the amplitudes have the asymptotic values
\beq
a\sup{n}_k=\frac{(k+n)!}{(n!)^2}.
\label{lead}
\eeq
We thus obtain in particular
\beq
\mean{(Z\Z)^k}\approx\frac{k!}{D^k}.
\eeq
We read off from this result that the positive variable $S=R^2=Z\Z=X^2+Y^2$
is asymptotically exponentially distributed at large $D$,
with density $f_S(s)\approx D\e^{-Ds}$.
In other words, $Z$ is asymptotically a complex Gaussian variable,
with the isotropic density
\beq
f(x,y)\approx\frac{D}{\pi}\,\e^{-DR^2}.
\label{bulk}
\eeq

The analysis of the recursion~(\ref{leadrec}) can be pursued to derive
a systematic~$1/D$-expansion beyond the asymptotic values~(\ref{lead}).
For the diagonal moments, we thus obtain after some algebra
\beq
\mean{(Z\Z)^k}=\frac{k!}{D^k}\left(1-\frac{k(7k+1)}{8D}+\cdots\right).
\eeq

\subsection{Qualitative features}
\label{dep}

In this section we make a pause in our analytical investigations
and turn to a qualitative discussion of various features of the distribution
of the variable $Z$, supported by numerical results.

\begin{figure}[!ht]
\begin{center}
\includegraphics[angle=-90,width=.3\linewidth]{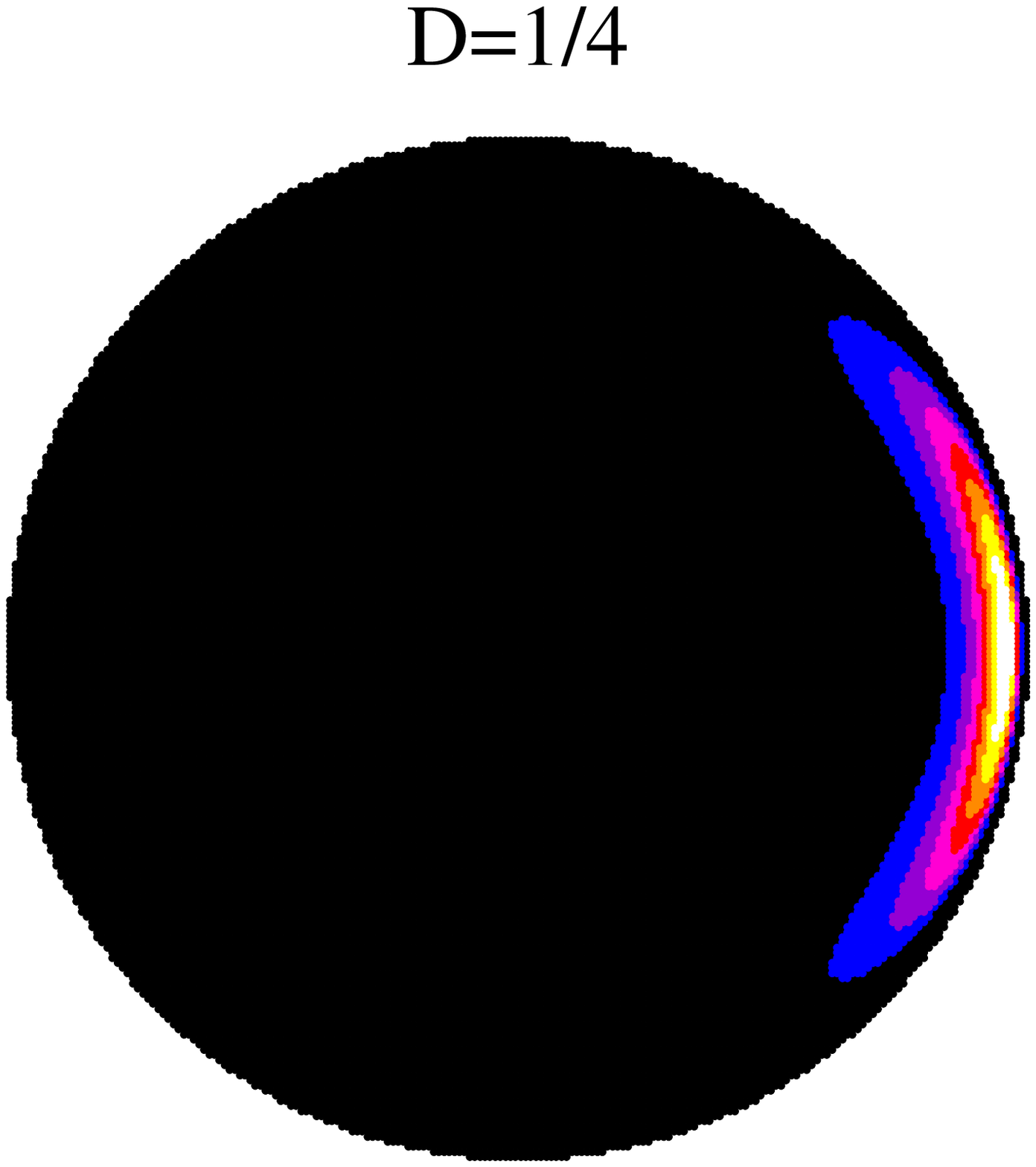}
\includegraphics[angle=-90,width=.3\linewidth]{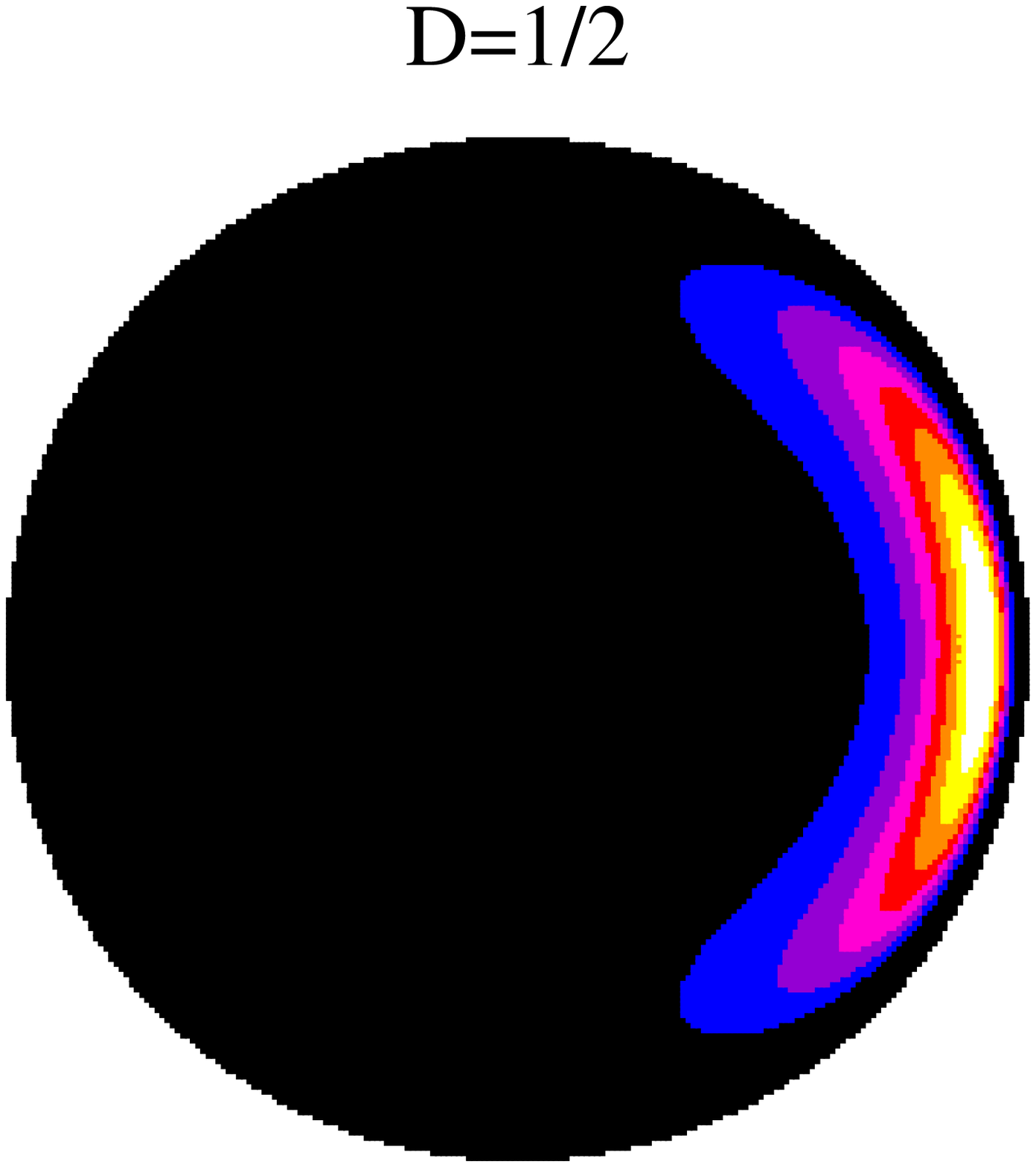}
\includegraphics[angle=-90,width=.3\linewidth]{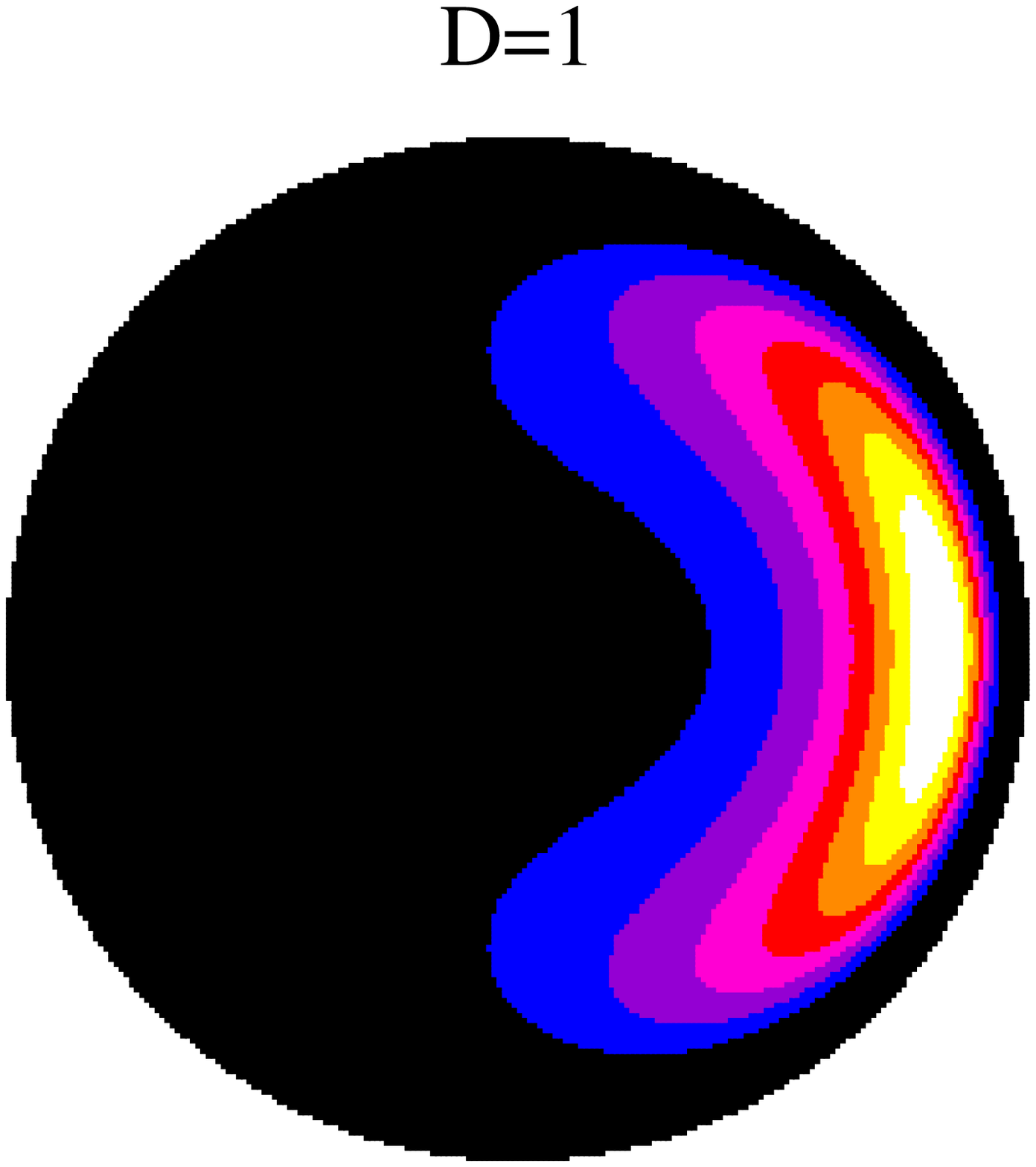}
\vskip 14pt
\includegraphics[angle=-90,width=.3\linewidth]{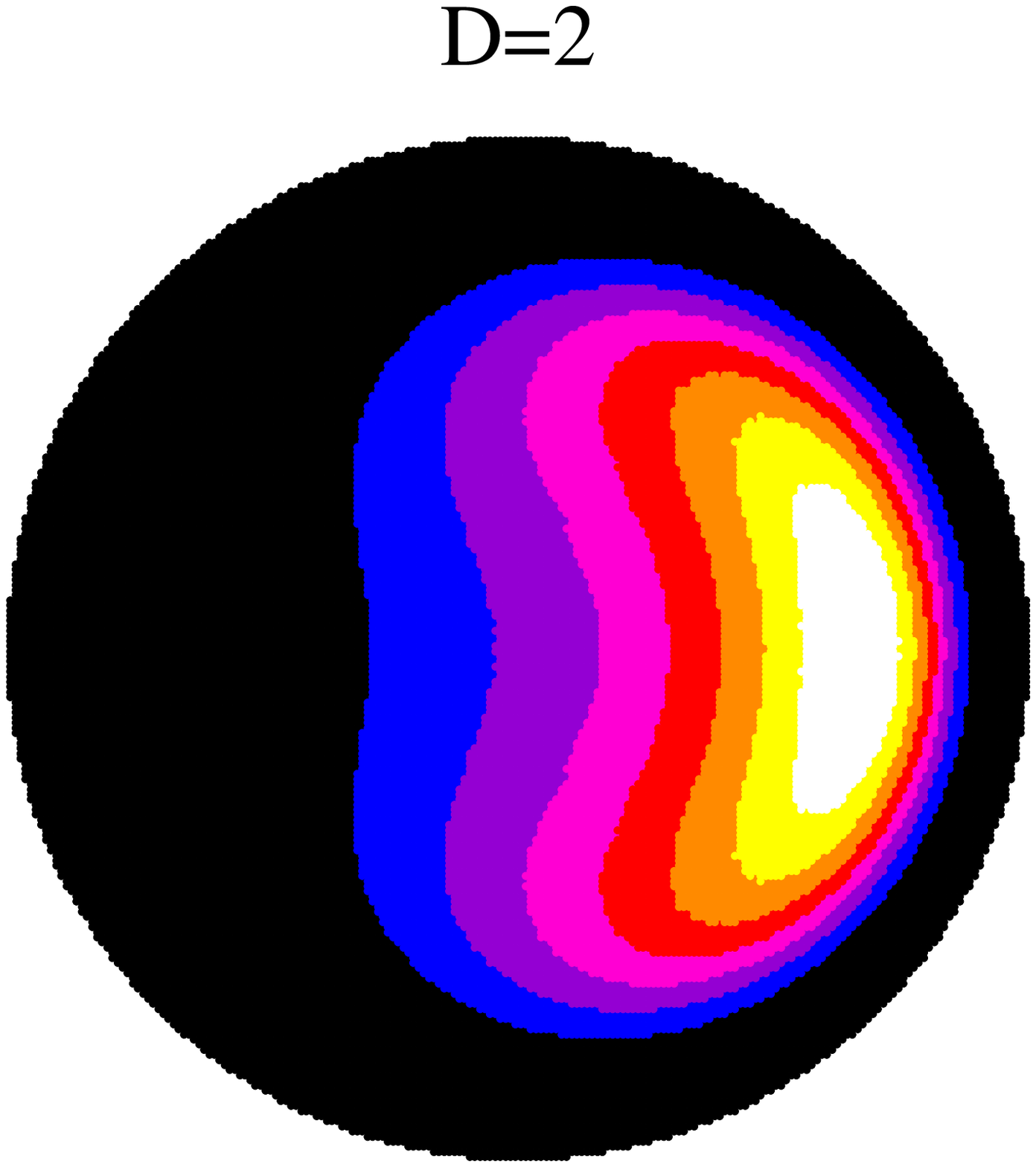}
\includegraphics[angle=-90,width=.3\linewidth]{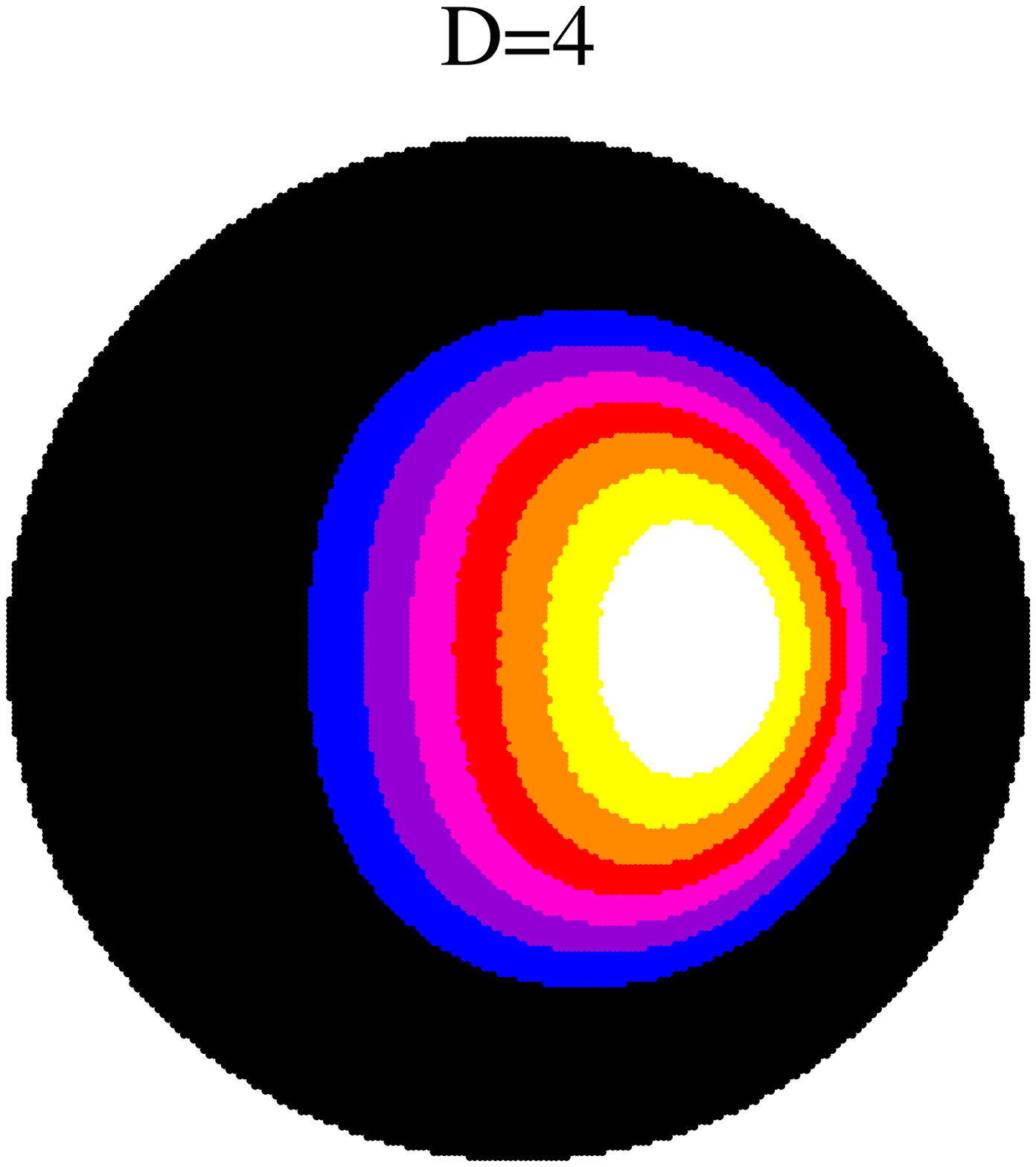}
\includegraphics[angle=-90,width=.3\linewidth]{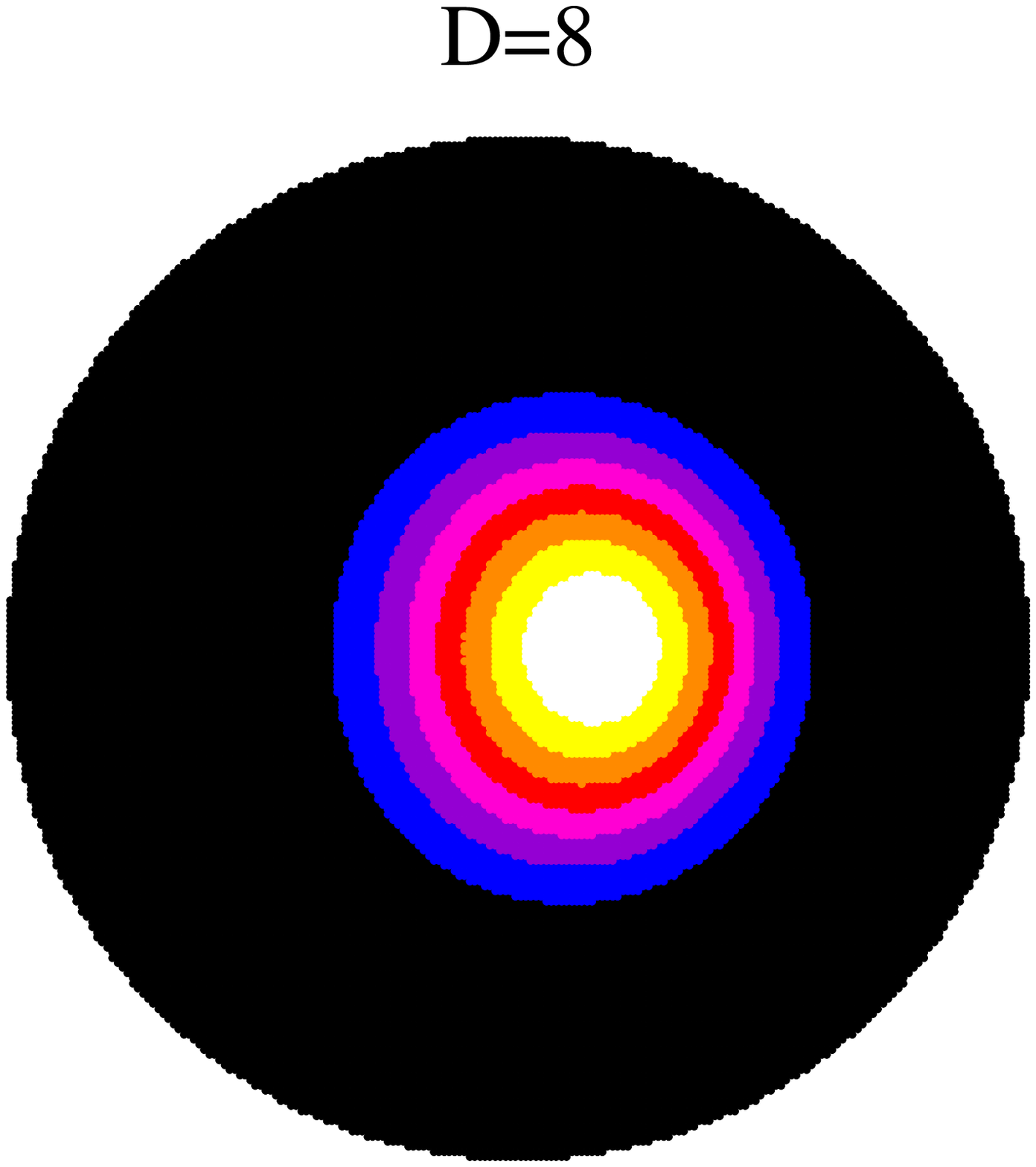}
\caption{\label{plots}
Level plots of the density $f(x,y)$
of the variable $Z=X+\i Y$ in the unit disk,
for several values of $D$.}
\end{center}
\end{figure}

\begin{figure}[!ht]
\begin{center}
\includegraphics[angle=-90,width=.45\linewidth]{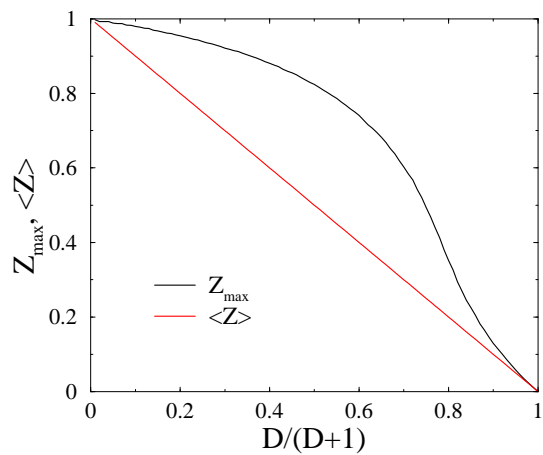}
\caption{\label{zmax}
Plot of the location $Z_\max$ of the maximum of the distribution of $Z$
and of the mean $\mean{Z}$ (see~(\ref{zmean})), against $D/(D+1)$.}
\end{center}
\end{figure}

The global picture is provided by Figure~\ref{plots},
showing color level plots of the density $f(x,y)$
of the variable $Z=X+\i Y$ in the unit disk,
for several values of $D$.
The data presented here and throughout the following
have been obtained by numerically generating very long time series
of the discrete process~(\ref{zrec1}),
with typically $\eps=10^{-3}$ and $n=10^{10}$.
First of all, as the distribution is symmetric w.r.t.~the $X$-axis,
its unique maximum is on the real axis.
Figure~\ref{zmax} shows a plot of the observed location $Z_\max$
of this maximum (black curve)
and of the mean $\mean{Z}$ (red line, see~(\ref{zmean})), against $D/(D+1)$.
Both quantities qualitatively follow the same pattern,
decreasing from 1 in the $D\to0$ limit to 0 in the $D\to\infty$ limit.

The full shape of the distribution is observed to exhibit
a smooth dependence on the diffusion constant $D$,
continuously interpolating between the weak-disorder and strong-disorder regimes,
respectively studied in Sections~\ref{weak} and~\ref{strong}.

\begin{itemize}

\item
For small $D$, the distribution of $Z$
is concentrated in a very elongated domain
which is stretched along the unit circle and fits its curvature.
This very anisotropic distribution
is best illustrated on the first panel of Figure~\ref{plots}, corresponding to $D=1/4$.
The picture corroborates the fact
that $Y$ has a variance $\mean{Y^2}\approx D$,
and therefore a vertical extent scaling as~$\sqrt{D}$,
whereas the horizontal extent of $1-X\approx DV$ is much smaller.
The curvature reflects the property
that $U^2$ and $V$ are strongly positively correlated.

\item
For large $D$, $Z$ is predicted to be asymptotically a complex Gaussian variable
such that $\mean{R^2}\approx1/D$.
Its distribution is therefore essentially concentrated
in a small and nearly circular domain near the origin.
This nearly isotropic distribution is illustrated on the last panel of
Figure~\ref{plots}, corresponding to $D=8$.

\end{itemize}

\begin{figure}[!ht]
\begin{center}
\includegraphics[angle=-90,width=.45\linewidth]{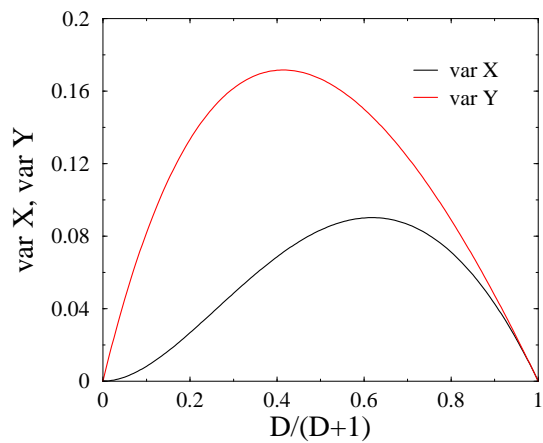}
\caption{\label{vars}
Plot of the variances $\var X$ and $\var Y$ (see~(\ref{varsres}))
against $D/(D+1)$.}
\end{center}
\end{figure}

The expressions of the variances of $X$ and $Y$,
\beq
\var X=\frac{D^2}{(D+1)^2(2D+1)},\qquad\var Y=\frac{D}{(D+1)(2D+1)},
\label{varsres}
\eeq
are readily derived from Table~\ref{momtab}.
Both variances go to zero at both endpoints ($D\to0$ and $D\to\infty$).
For small $D$, $\var X\approx D^2$ vanishes much faster than $\var Y\approx D$,
reflecting the very anisotropic nature of the distribution.
To the contrary, at large $D$
both variances are asymptotically equal to each other,
as $\var X\approx\var Y\approx1/(2D)$,
reflecting the nearly isotropic nature of the distribution.
The variances are maximal for intermediate values of the diffusion constant.
The maximum $(\var X)_\max=(5\sqrt{5}-11)/2\approx0.090169$ of $\var X$
is reached for $D=(1+\sqrt{5})/2\approx1.618033$,
whereas $(\var Y)_\max=3-2\sqrt{2}\approx0.171572$
is reached for $D=1/\sqrt{2}\approx0.707106$.
It is worth noticing that $\var Y$ is invariant if $D$ is changed to
its dual $\w D$, such that
\beq
D\w D=\frac{1}{2}.
\label{dual}
\eeq
The maximum $(\var Y)_\max$ is consistently reached at the fixed point
$D=1/\sqrt{2}$ of this duality transformation.
Figure~\ref{vars} shows a plot of both variances against $D/(D+1)$.

\begin{figure}[!ht]
\begin{center}
\includegraphics[angle=-90,width=.45\linewidth]{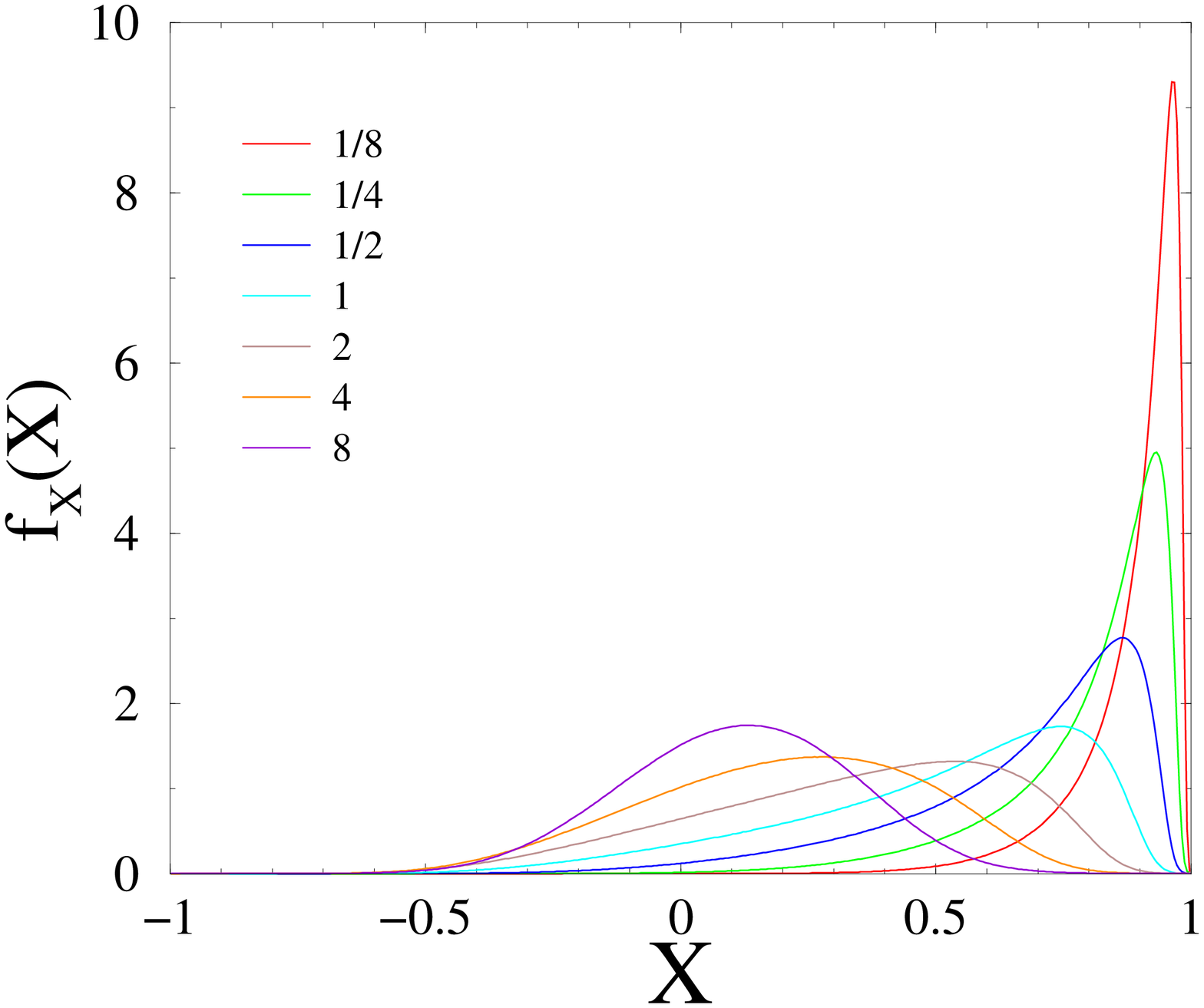}
\qquad
\includegraphics[angle=-90,width=.45\linewidth]{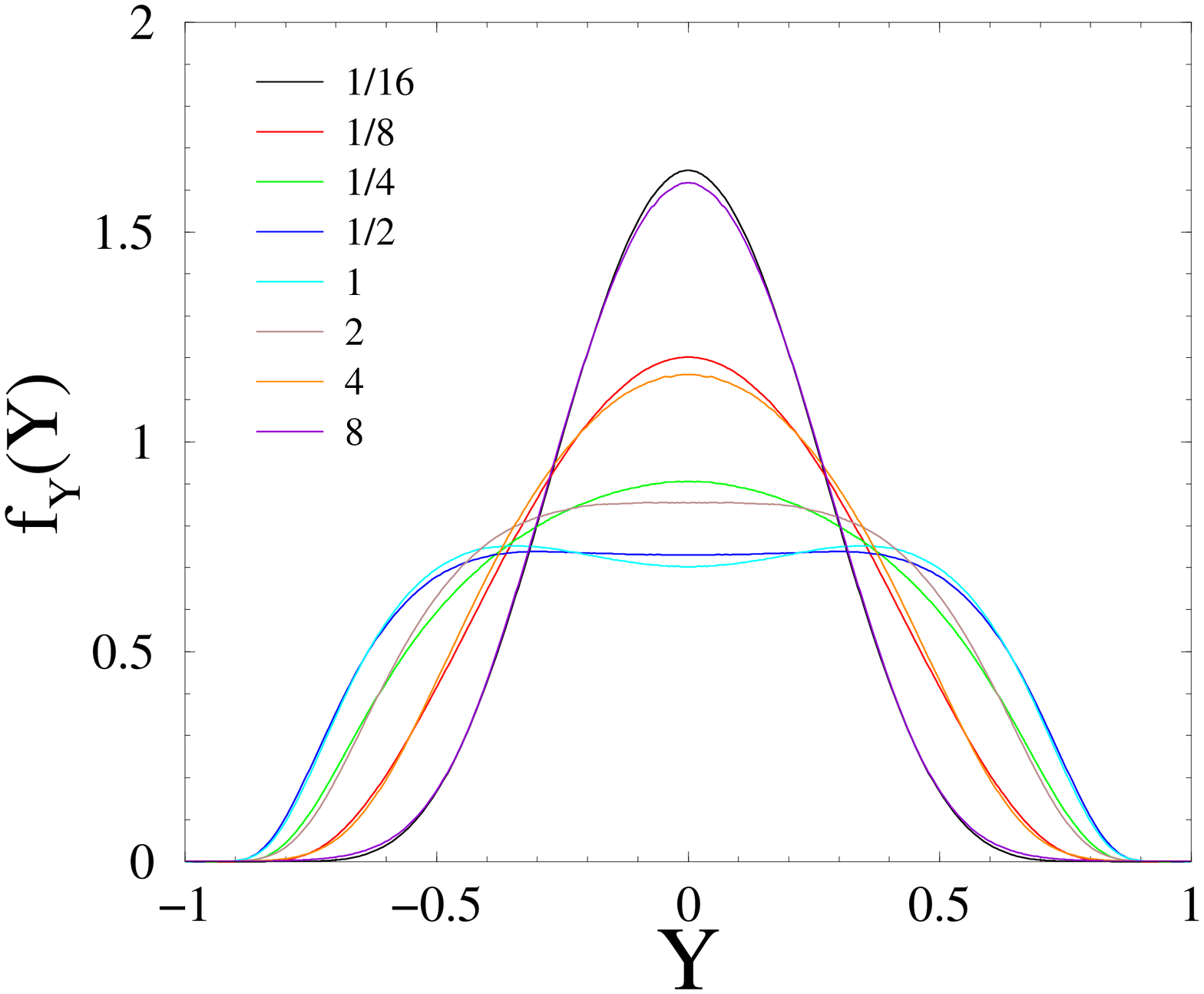}
\caption{\label{histos}
Plot of the densities $f_X$ (left) and $f_Y$ (right)
of the marginal distributions of $X$ and $Y$, for values of $D$
chosen so as to form the dual pairs (8, 1/16), (4, 1/8), (2, 1/4), and (1, 1/2)
(see~(\ref{dual})).}
\end{center}
\end{figure}

The features discussed so far are corroborated
by the shape of the marginal distributions $f_X$ and $f_Y$
of the real variables $X$ and $Y$.
These distributions, obtained as numerical histograms,
are shown in Figure~\ref{histos} for several values of $D$,
chosen so as to form dual pairs (see~(\ref{dual})).
The dependence of the density of $X$ on the diffusion constant
is observed to evolve smoothly from a narrow distribution around $X=1$ at small $D$,
whose limiting shape is given by the density $f_V$ of the rescaled variable~$V$,
to a narrow Gaussian near $X=0$ at large $D$,
whose variance scales as $\var X\approx1/(2D)$.
The distribution of $Y$ behaves differently.
It indeed becomes a narrow Gaussian both at weak and at strong disorder,
whose variance respectively falls off as $\var Y\approx D$ ($D\to0$)
and $\var Y\approx1/(2D)$ ($D\to\infty$).
The densities of $Y$ for pairs of dual values $D$ and $\w D$
of the diffusion constant are observed to be strikingly close to each other.

\subsection{Asymptotic behavior near the boundary: an essential singularity}
\label{circle}

We now resume our analytical investigations and turn to more advanced features,
namely the asymptotic analysis of the distribution of $Z$
in two regimes where it is exponentially small:
near the boundary (unit circle) for all values of $D$ (in this section),
and for large~$D$ at any point inside the unit disk (in Section~\ref{larged}).

For the time being,
our goal is to investigate the distribution of $Z$ near the unit circle.
More precisely, we are interested in the behavior of the density $f_\nu(X_\nu)$
of the variable $X_\nu$ defined in~(\ref{xnudef})
as the unit circle is approached from inside ($X_\nu\to1$),
for a fixed value of the angle $\nu$.
The definition~(\ref{omegadef}) implies
\beq
f_\nu(X_\nu)=\int\frac{\d\mu}{2\pi\i}\,\e^{-\mu X_\nu}\,\Omega(\mu,\nu).
\label{fnudef}
\eeq
We are thus led to study the asymptotic behavior
of the characteristic function $\Omega(\mu,\nu)$ at large $\mu$ and fixed $\nu$.

It is worthwhile to first have a look at the simplified situation considered
in Section~\ref{weak}, i.e., $\nu\ll1$.
In this regime, inserting~(\ref{absol}) into~(\ref{simple}),
and expanding the result for large $\mu$, we obtain
\beqa
\Omega(\mu,\nu)
=\exp\,\Biggl\{\mu&-&\sqrt{2D\mu}\left(1+\frac{\nu^2}{4D}\right)\nonumber\\
&+&\frac{1}{8}\ln(32\pi^2D\mu)-\frac{\nu^2}{16D}+\cdots\,\Biggr\}.
\label{simpleasy}
\eeqa

Inspired by the above expansion,
we look for a solution of the full partial differential equation~(\ref{munupde})
which behaves at large $\mu$ as
\beq
\Omega(\mu,\nu)=\exp\Bigl(\mu-A(\nu)\sqrt{\mu}+\cdots\Bigr).
\label{asy}
\eeq
Inserting this expansion into~(\ref{munupde}),
we are left with the ordinary differential equation
\beq
\left(\frac{\d A}{\d\nu}\right)^2=\frac{1-\cos\nu}{D}=\frac{2}{D}\sin^2\frac{\nu}{2}
\eeq
for the amplitude $A(\nu)$.
Let us assume that the value $A(0)=\sqrt{2D}$ at $\nu=0$,
which can be read off from~(\ref{simpleasy}), holds for all values of $D$.
This hypothesis will be tested against numerical results in a while.
We thus obtain
\beq
A(\nu)=\sqrt{\frac{2}{D}}\left(D+2-2\cos\frac{\nu}{2}\right).
\label{ares}
\eeq

The fall-off of the distribution of $Z$ near the unit circle
can be estimated by inserting~(\ref{asy}) into~(\ref{fnudef}):
\beq
f_\nu(X_\nu)\sim\int\frac{\d\mu}{2\pi\i}\,\e^{\mu(1-X_\nu)-A(\nu)\sqrt{\mu}}.
\eeq
For $X_\nu\to1$, this integral is dominated by a saddle point
at $\mu\approx A(\nu)^2/(4(1-X_\nu)^2)$.
This yields the estimate
\beq
f_\nu(X_\nu)\sim\exp\left(-\frac{A(\nu)^2}{4(1-X_\nu)}\right),
\eeq
i.e.,
\beq
f_\nu(X_\nu)
\sim\exp\left(-\frac{\left(D+2-2\cos(\nu/2)\right)^2}{2D(1-X_\nu)}\right).
\label{fnuexp}
\eeq

This result confirms that the distribution of $Z$ is supported by the whole unit disk,
and predicts that it vanishes exponentially fast near any point of the unit circle.
The behavior of the densities of the variables $X$ and $Y$ near their endpoints
is given by the following simpler formulas,
respectively corresponding to $\nu=0$, $\nu=\pm\pi$, and $\nu=\pm\pi/2$:
\beq
\matrix{
f_X(x)\sim\exp\left(-\frad{D}{2(1-x)}\right)\hfill&(x\to1),\hfill\cr
f_X(x)\sim\exp\left(-\frad{(D+2)^2}{2D(1+x)}\right)\hfill&(x\to-1),\hfill\cr
f_Y(y)\sim\exp\left(-\frad{(D+2-\sqrt{2})^2}{2D(1-\abs{y})}\right)\qquad&(y\to\pm1).\hfill}
\eeq
The first of these estimates matches the behavior~(\ref{fvinf}) of the density of $V$.

The dependence of the amplitude $A(\nu)$ on $D$ is not monotonic.
It diverges at both endpoints ($D\to0$ and $D\to\infty$),
and therefore reaches a minimum, $A_\min(\nu)=4\sqrt{1-\cos(\nu/2)}$,
for an angle-dependent value of the diffusion constant,
$D_\min(\nu)=2-2\cos(\nu/2)$.
This non-monotonic behavior holds for any angle except along the real axis ($\nu=0$),
where we have $A(0)=\sqrt{2D}$.
It is reminiscent of the behavior of $\var Y$ discussed in Section~\ref{dep}.

\begin{figure}[!ht]
\begin{center}
\includegraphics[angle=-90,width=.45\linewidth]{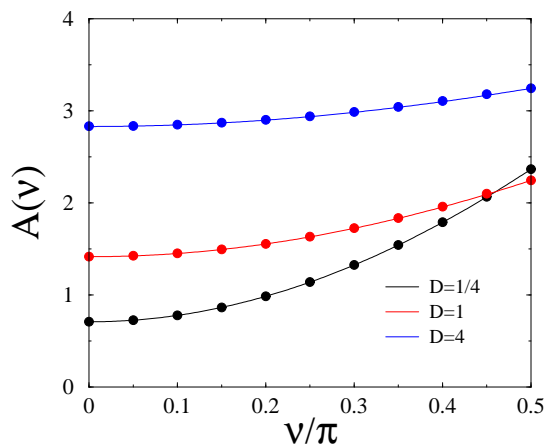}
\caption{\label{anu}
Plot of the amplitude $A(\nu)$ against the reduced angle $\nu/\pi$,
for three values of the diffusion constant $D$.
Full lines: analytical prediction~(\ref{ares}).
Symbols: result of a fit of the first 1,000 polynomials $P_k(\cos\nu)$
(see text).}
\end{center}
\end{figure}

The essential singularity~(\ref{fnuexp})
of the distribution of $Z$ near the unit circle is unobservable in numerical data.
Nevertheless, it has an observable consequence.
It indeed reflects itself in the large-order behavior of the polynomials
$P_k(\cos\nu)=\mean{X_\nu^k}$ (see~(\ref{pkdef})).
We have
\beq
P_k(\cos\nu)=\int_{-1}^1X_\nu^k\,f_\nu(X_\nu)\,\d X_\nu.
\eeq
For large $k$, the integral is dominated by either of the endpoints.
For an angle $\nu<\pi/2$, the leading endpoint is $X_\nu\to1$,
so that the integral can be estimated as
\beq
P_k(\cos\nu)\sim\int_{-1}^1X_\nu^k\exp\left(-\frac{A(\nu)^2}{4(1-X_\nu)}\right)\,\d X_\nu.
\eeq
The integral is dominated by a saddle point at $X_\nu\approx1-A(\nu)/(2\sqrt{k})$, yielding
\beq
P_k(\cos\nu)\sim\exp\left(-A(\nu)\sqrt{k}\right).
\label{xexp}
\eeq
We therefore predict a stretched exponential decay of the polynomials $P_k(\cos\nu)$,
at any fixed angle $\nu$.
Such a behavior can easily be observed numerically.
The recursion~(\ref{momrec}) and the expansion~(\ref{pkexpand})
indeed allow a very accurate numerical evaluation of at least
the first 1,000 polynomials.
Figure~\ref{anu} shows a plot of the amplitude $A(\nu)$
against the reduced angle $\nu/\pi\le1/2$,
for several values of the diffusion constant $D$.
The striking agreement between the prediction~(\ref{ares}) for the amplitude
(full lines)
and the result of a quadratic fit of {\it all} the first 1,000 polynomials
as $\ln P_k(\cos\nu)=-A\sqrt{k}+B\ln k+C$ (symbols)
shows in particular that the hypothesis made above~(\ref{ares}) is valid.

\subsection{Large deviations at large $D$}
\label{larged}

We close our investigation
by returning to the regime where the diffusion constant $D$ is large.
In this situation, it has been shown in Section~\ref{strong}
that the bulk of the distribution of $Z$ is a narrow isotropic Gaussian,
whose density is proportional to $\e^{-DR^2}$ (see~(\ref{bulk})).
Furthermore, the estimate~(\ref{fnuexp}) of the distribution
near the unit circle also becomes isotropic at large $D$,
and it falls off at $\e^{-D/(2(1-R))}$.
These observations suggest that the distribution follows
an exponential large-deviation estimate of the form
\beq
f(x,y)\approx f(R)\sim\e^{-DS(R)}
\label{sdef}
\eeq
for large $D$, all over the unit disk.

Our goal is to show that such an estimate indeed holds,
and to derive the large-deviation function $S(R)$.
First, if the distribution of $Z$ is nearly isotropic, we have
\beq
\Omega(\mu,\nu)\approx2\pi\int_0^\infty Rf(R) I_0(\mu R)\,\d R,
\eeq
for all values of $\nu$.
Then, if the distribution obeys the formula~(\ref{sdef}),
the above integral can be evaluated for large $\mu$
(i.e., $\mu$ of the order of $D$)
by means of the saddle-point approximation.
Setting $\mu=Dx$, we obtain
\beq
\Omega(\mu,\nu)\sim\e^{D\S(x)},
\label{sigdef}
\eeq
where the functions $S(R)$ and $\S(x)$ are the Legendre transforms of each other:
\beq
\S(x)+S(R)=xR,\qquad x=S'(R),\qquad R=\S'(x).
\eeq
The formulas~(\ref{sdef}) and~(\ref{sigdef}) are just exponential estimates.
In order to proceed, we have to consider the angular dependence
of the pre-exponential factor in the latter estimate.
Setting
\beq
\Omega(\mu,\nu)\approx H(\nu)\,\e^{D\S(x)},
\eeq
the partial differential equation~(\ref{munupde})
yields the ordinary differential equation
\beq
H''(\nu)+(x\cos\nu-E)H(\nu)=0
\label{madef}
\eeq
for the amplitude $H(\nu)$ at fixed $x=\mu/D$, with $E=xR=\mu R/D$.
Equation~(\ref{madef}) is known as the Mathieu equation~\cite{gr}.
The requirement that $H(\nu)$ be periodic and positive
implies that $E(x)$ is the ground-state eigenvalue of the latter equation.
Introducing the Fourier~series
\beq
H(\nu)=\sum_{n=-\infty}^\infty h_n\,\e^{\i n\nu},
\eeq
the Mathieu equation~(\ref{madef}) is equivalent to the three-term recursion
\beq
2(n^2+E)h_n=x(h_{n+1}+h_{n-1}),
\eeq
which is the appropriately rescaled form of the recursion~(\ref{difdif}).
The ground-state eigenvalue $E(x)$ is characterized by the property
that the above recursion has a positive normalizable solution.
The function $E(x)$ being known,
$\S(x)$ is obtained by integrating the relation $\S'(x)=E(x)/x$ with $\S(0)=0$,
and finally we have $R=E(x)/x$ and $S(R)=E(x)-\S(x)$.
We have thus obtained the large-deviation function $S(R)$ in parametric form.
Analytical expansions can be performed for $x\to0$ and $x\to\infty$,
respectively corresponding to $R\to0$ and $R\to1$.

For $x\to0$, the amplitudes $h_n\sim x^n$ of the harmonics fall off very fast.
Taking successive harmonics into account, we can derive the expansions
\beqa
E(x)&=&\frac{1}{2}\,x^2-\frac{7}{32}\,x^4+\frac{29}{144}\,x^6+\cdots\qquad(x\to0),\nonumber\\
S(R)&=&R^2+\frac{7}{8}\,R^4+\frac{395}{432}\,R^6+\cdots{\hskip 34.25pt}(R\to0).
\eeqa

For $x\to\infty$, the function $H(\nu)$ becomes peaked around $\nu=0$.
In this regime,~(\ref{madef}) maps onto a weakly anharmonic quantum-mechanical oscillator.
Treating the leading non-linearity in $\nu^4$ to first order, we obtain
\beqa
E(x)&=&x-\sqrt{\frac{x}{2}}+\frac{1}{16}+\cdots{\hskip 102pt}(x\to\infty),\nonumber\\
S(R)&=&\frac{1}{2(1-R)}+\frac{1}{8}\,\ln(1-R)+C+\cdots\qquad(R\to1).
\eeqa

The leading behavior of $S(R)$ at both endpoints matches
the known results recalled in the beginning of this section.
The coefficient $1/8$ of the logarithmic term in $S(R)$ near $R=1$
matches the similar coefficient in~(\ref{simpleasy}).

\begin{figure}[!ht]
\begin{center}
\includegraphics[angle=-90,width=.43\linewidth]{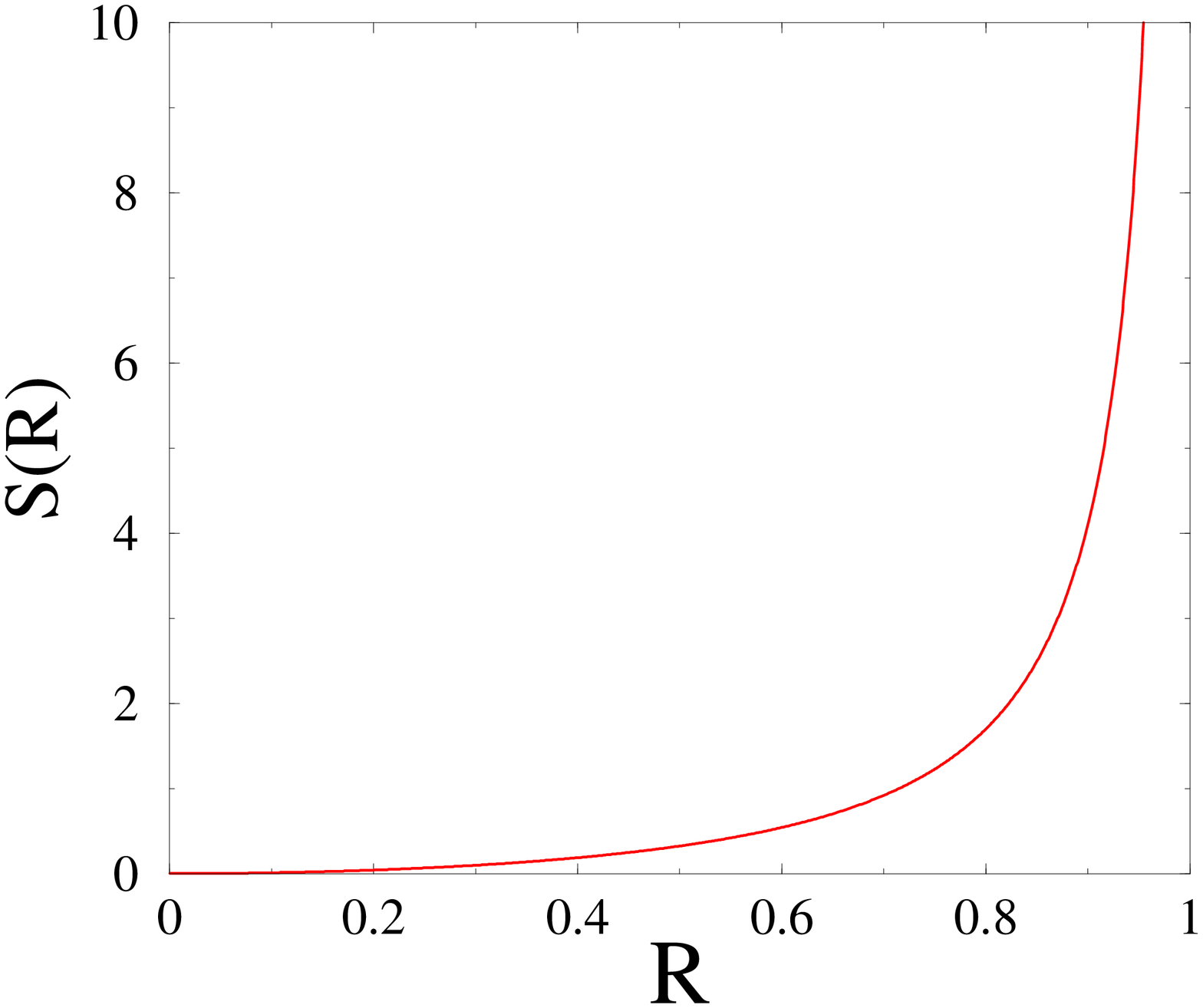}
\qquad
\includegraphics[angle=-90,width=.45\linewidth]{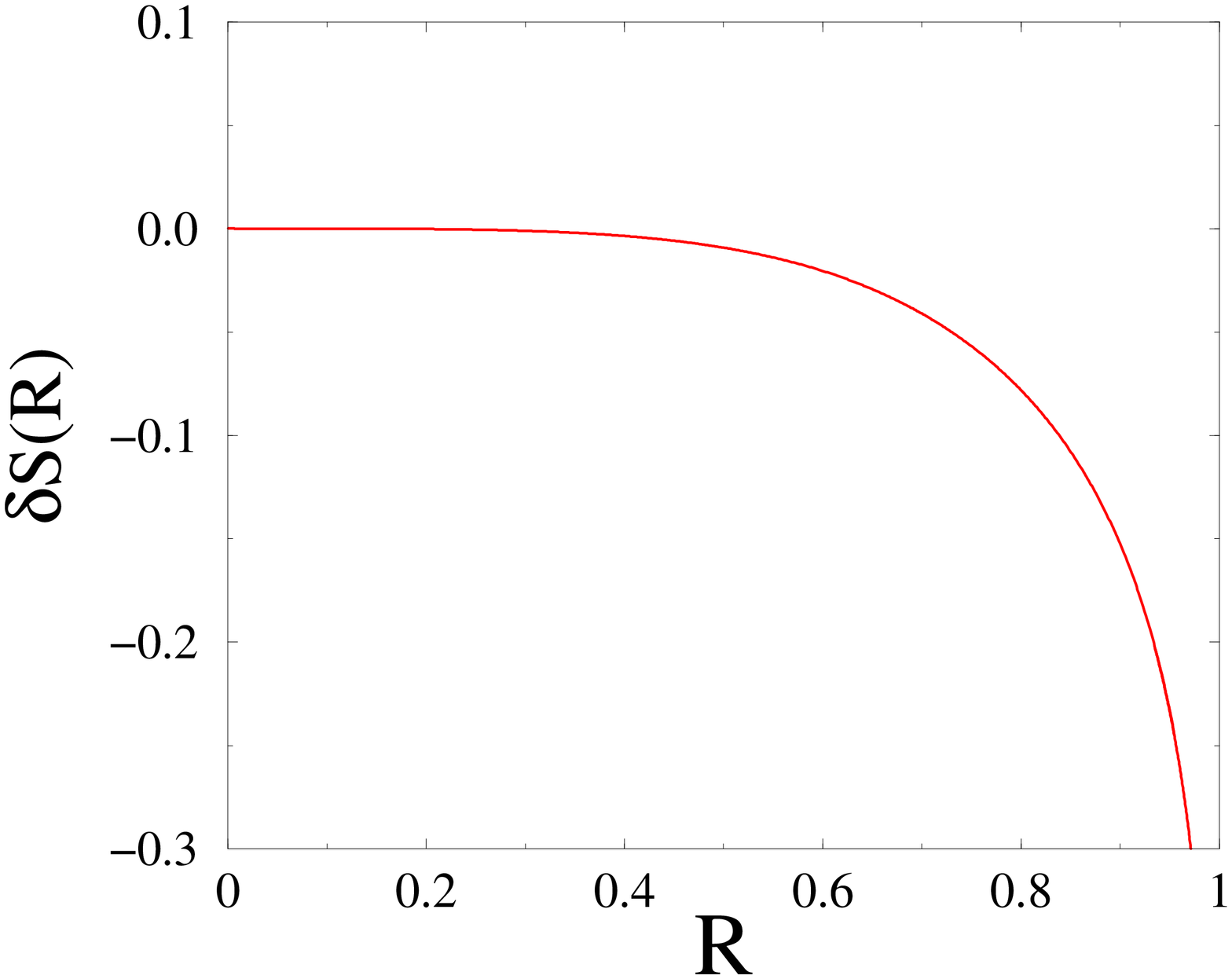}
\caption{\label{sr}
Plot of the large-deviation function $S(R)$ characterizing
the distribution of~$Z$ at large $D$ (left),
and of the difference $\delta S(R)=S(R)-R^2/(1-R^2)$ (right).}
\end{center}
\end{figure}

The above scheme is easily implemented numerically.
This yields e.g.~the constant $C\approx-0.603$.
Figure~\ref{sr} shows a plot of the function $S(R)$ thus obtained (left),
and of the small difference $\delta S(R)=S(R)-R^2/(1-R^2)$ (right).

\section{Discussion}
\label{discussion}

This work has been devoted to a detailed investigation
of a random complex integral~$Z$,
which provides the most natural example
of an imaginary exponential functional of Brownian motion.
Both in the warming-up part (Section~\ref{real}) devoted to the real functional $X$
and in the main part (Section~\ref{imag}) devoted to $Z$,
the main emphasis has been put on the complementarity between
the more traditional approach, based on Langevin equations and stochastic calculus,
and a more original one, based on discrete random recursions and Kesten variables.
Even though neither of these routes leads to the solution of the problem,
i.e., to an explicit expression for the probability distribution function of $Z$,
we have gathered many partial results concerning various facets
of this distribution in various regimes.
Let us emphasize in particular the recursion relation~(\ref{momrec})
for the moments $\mean{Z^k\Z^l}$,
and the occurrence of an essential singularity of the form~(\ref{fnuexp})
of the asymptotic behavior of the density near the unit circle.

Apart from its intrinsic interest,
the imaginary exponential functional $Z$
has close connections with the realm of reaction-diffusion processes.
This fact, already underlined in~\cite{delou},
was our original motivation for pursuing the present work.
The ubiquity of {\it imaginary noise} Langevin equations
illustrates the potential pitfalls one faces when building up
phenomenological stochastic equations in out-of-equilibrium situations.
In fact, within the Doi-Peliti formalism for reaction-diffusion processes,
it can be shown~\cite{TBP} that degenerate Langevin equations for a complex-valued field
are the rule rather than the exception.
The present study will hopefully trigger further interest in such stochastic processes.

\ack

We warmly thank Robert Conte for many discussions and for his encouragements.

\end{document}